\documentclass[aip, amsmath, amssymb, reprint,pop,floatfix]{revtex4-2}
\usepackage{amsfonts}
\usepackage[utf8]{inputenc}
\usepackage{graphicx} 

\usepackage[hidelinks]{hyperref}
\usepackage{xcolor}
\newcommand*{\citen}[1]{%
  \begingroup
    \romannumeral-`\x 
    \setcitestyle{numbers}%
    \cite{#1}%
  \endgroup   
}
\newcommand{\pdv}[2]{\frac{\partial #1}{\partial #2}}
\newcommand{\dv}[2]{\frac{\mathrm{d} #1}{\mathrm{d} #2}}

\begin{document}

\title{Magnetostatic Ponderomotive Potential in Rotating Plasma} 

\author{T. Rubin}
\email{trubin@princeton.edu}
\affiliation{Department of Astrophysical Sciences, Princeton University, Princeton, New Jersey 08540, USA}

\author{J. M. Rax}
\affiliation{Andlinger Center for Energy + the Environment, Princeton University, Princeton, New Jersey 08540, USA}
\affiliation{IJCLab, Universit\'{e} de Paris-Saclay, 91405 Orsay, France}

\author{N. J. Fisch}
\affiliation{Department of Astrophysical Sciences, Princeton University, Princeton, New Jersey 08540, USA}
\date{\today}
\begin{abstract}
	{A new end-plugging method for rotating plasmas is identified and analyzed. It uses the ponderomotive potential associated with an azimuthal magnetostatic wiggler. Studied both analytically and numerically, this process compares favorably to other end-plugging methods in open field line magnetized plasma devices. } 
\end{abstract}

\maketitle
\section{Introduction}

Magnetic confinement of plasma relies on the basic behavior of charged particles in static homogeneous magnetic fields: the orbit is the combination of (\textit{i}) a rotation around the field lines and (\textit{ii}) a translation along the field lines, with both rotation and translation uniform. Given this basic behavior, to design a magnetic trap with an inhomogeneous static magnetic field, two types of configurations can be considered:

(\textit{i}) Open field lines configurations, where the magnetic field lines are closed outside the plasma. The configuration must display a minimum of a trapping potential along the open field lines to restrict the parallel motion and to achieve confinement. For nonneutral plasma in a Penning trap this potential is electrostatic, and for thermonuclear quasineutral plasma this potential is associated with the diamagnetic force leading to magnetic  mirror confinement\cite{postMagneticMirrorApproach1987,ryutovOpenendedTraps1988}.

(\textit{ii}) Closed field lines configurations, associated with a toroidal topology, are also well suited for confinement. There is no need to create a minimum of a potential along the field lines as particles explore the full length of the line. But the magnetic toroidal vertical drift velocity across the field lines is to be compensated. This compensation is achieved with the \textit{rotational transform}\cite{spitzerStellaratorConcept1958,budkerPlasmaPhysicsProblems1959} which allows a short circuiting of the vertical drift currents, thus providing steady state confinement. Stellarators and tokamaks are the two main configurations designed according to this principle.

These two magnetic confinement principles: (\textit{i}) open field lines together with the necessity of end-plugging field lines  to avoid the parallel escape along the lines; and (\textit{ii}) closed field lines together with the necessity of a rotational transform to compensate the vertical drift escape, have been considered since the early times of thermonuclear plasma physics.

In open traps of the mirror type, despite the occurrence of a minimum of the diamagnetic potential, some particles with large parallel velocity escape the confined plasma and additional confining forces are to be considered. Several schemes have been identified to provide additional end-plugging of classical mirrors configurations. Two classic review papers [\citen{gormezanoReductionLossesOpenended1979,baldwinEndlossProcessesMirror1977}] summarize the principles and experimental achievements related to end-plugging in mirror traps. Among the different principles identified to stop the escaping particles, the ponderomotive force associated with Radio Frequency (RF) waves offers a straightforward scheme. However, to maintain end plugging through ponderomotive potentials at the reactor scale is power intensive.
Confinement in mirror traps can also be enhanced by the centrifugal force\cite{lehnertRotatingPlasmas1971,bekhtenevProblemsThermonuclearReactor1980,volosovLongitudinalPlasmaConfinement1981,hassamSteadyStateCentrifugally1997,teodorescuConfinementPlasmaShaped2010,fettermanAlphaChannelingRotating2010,fettermanWavedrivenRotationSupersonically2010,fettermanChannelingRotatingPlasma2008,fowlerNewSimplerWay2017,whiteCentrifugalParticleConfinement2018}.The centrifugal force, acting mainly on ions,  results from a rotation sustained by an electric field perpendicular to the magnetic surfaces. Another loss-reduction scheme combining plasma rotation and magnetostatic features has been explored, where the mirror throat is twisted into a helix\cite{beklemishevHelicoidalSystemAxial2013,beklemishevHelicalPlasmaThruster2015,postupaevHelicalMirrorsActive2016}. One can also use plasma rotation and electrostatic perturbations\cite{andereggLongIonPlasma1995} rather than magnetostatic ones, which might be generated by external azimuthal structures to assist in plasma confinement.

 The end-plugging of a rotating plasma column has utility not only for enhancing the mirror confinement of rotating hot plasmas for the purpose of controlled nuclear fusion, but also has uses in other applications such as: 
 (\textit{i}) mass separation with rotating
plasmas envisioned for nuclear waste cleanup and spent nuclear fuel reprocessing;\cite{gueroultPlasmaFilteringTechniques2015,dolgolenkoSeparationMixturesChemical2017,timofeevTheoryPlasmaProcessing2014,gueroultPlasmaMassFiltering2014,voronaPossibilityReprocessingSpent2015,litvakArchimedesPlasmaMass2003}  (\textit{ii}) $\mathbf{E}\times\mathbf{B}$ plasma configurations for the purpose of ions acceleration;\cite{janesExperimentsMagneticallyProduced1965,janesNewTypeAccelerator1966,janesProductionBeVPotential1965} and (\textit{iii}) thermonuclear fusion based upon rotation in toroidal confinement devices.\cite{raxEfficiencyWavedrivenRigid2017,ochsParticleOrbitsForcebalanced2017}

What is identified here is a new mechanism for rotating mirror end-plugging: the use of a simple annular static magnetic wiggler. 
By means of a  Hamiltonian analysis of single particle dynamics in a rotating plasma interacting with such an azimuthal wiggler, we show that this interaction results in an axial ponderomotive force. 
The occurrence of this end-plugging process is then confirmed numerically.
This means of end plugging provides confinement that can be in addition to the centrifugal and simple mirror confinement.

To compare to other methods of charged particle reflection, consider that a static magnetic field can reflect particles as can be seen in a classical magnetic mirror $B_{m}\left( z\right) $, the adiabatic equation of motion describing the $\mu \nabla B$ force is 
\begin{equation}
	\left. \frac{d\left\langle v_{z}\right\rangle }{dt}\right| _{\mu \nabla B}=-\frac{q^{2}}{4m^{2}}\rho^{2}\frac{dB_{m}^{2}}{dz}  \label{w87}
\end{equation}
where $\rho$ is the Larmor radius of the particle in the $B_{m}$ field.

We can also consider the standard RF plugging based on the ponderomotive force of an inhomogeneous electromagnetic wave with electric field amplitude $E\left( z\right) $, magnetic field amplitude $B_{w}\left( z\right) =k_{w}E/\omega_{w}$, with frequency $\omega _{w}$ and wave vector $k_{w}$.\cite{gormezanoReductionLossesOpenended1979,motzRadioFrequencyConfinementAcceleration1967, gaponov1958potential}

\begin{equation}
	\left. \frac{d\left\langle v_{z}\right\rangle }{dt}\right| _{wave}=-\frac{q^{2}}{4m^{2}\omega _{w}^{2}}\frac{dE^{2}}{dz}=-\frac{q^{2}}{4m^{2}}\frac{1}{k_{w}^{2}}\frac{dB_{w}^{2}}{dz}\label{class pond force}
\end{equation}
However, the sustainment of a standing wave structure through the matching between an antenna and the plasma dispersion relation is far less simple than the use of a magnetic mirror.

The forces in Eqs. (\ref{w87}) and (\ref{class pond force}) are both of the ponderomotive type, and rely on time scale separation between a fast and a slow motion. In the $\mu \nabla B$ case, conservation of the first adiabatic invariants relies on $v_z dB_m/dz \ll B \Omega_c$, while in the wave case $v_z dE/dz \ll E \omega_w$. The ponderomotive force we seek would rely on such adiabaticity condition. 
The ponderomotive potentials, $\Phi$, that relate to the ponderomotive forces by $F_z = - d\Phi/dz$ are independent of the length scales $L_z ^{-1} = d\ln B/dz$. In the case of non-adiabatic interaction, particles would experience a phase-dependent attractive or repulsive force, resulting in a quasilinear diffusion. 



An annular wiggler around the edge plasma column also offers the possibility of direct energy conversion of the high energy particles that escape the mirror along the field lines\cite{moirVenetianblindDirectEnergy1973,mileyFusionEnergyConversion1976,taniguchiStudiesChargeSeparation2010,takenoRecentAdvancementResearch2019,volosovRecuperationChargedParticle2005}, but to assess the possibility this combination of the magnetic, centrifugal and azimuthal wiggler forces requires a careful Hamiltonian analysis beyond the scope of this work.

Note that the plasma rotation offers the possibility of achieving what amounts to a conventional ponderomotive forces in the frame of reference of the rotating plasma, where charged particle see rapidly oscillating fields.  Ponderomotive barriers can be set up in non-rotating plasmas by imposing RF fields in a variety of physical contexts \cite{caryPonderomotiveForceLinear1977, dimonteEffectsNonadiabaticityApplications1983,konoPonderomotiveForceCyclotron1987, grossmanRFPonderomotiveForces1992, masuzakiReductionPlasmaHeat1995,tokman99, dodinPonderomotiveRatchetUniform2005, dodinQuantumlikeDynamicsClassical2005, dodinNonadiabaticPonderomotivePotentials2006,dodinNonadiabaticTunnelingPonderomotive2006}, including also one-way type walls for current drive applications  \cite{ suvorov88, litvakNonlinearEffectsElectron1993, fischCurrentDrivePonderomotive2003, dodinPonderomotiveBarrierMaxwell2004}.  However, all of these applications require RF waves to be set up within the plasma, which can be technologically demanding.

We  now proceed to study the dynamics of particles in a magnetized rotating plasma interacting with an azimuthal wiggler. 
This paper is organized as follows: In Sec. \ref{SII} we identify the ponderomotive force associated with an azimuthal wiggler within the framework of a Hamiltonian analysis. 
We confirm this result through direct numerical simulations in Sec.~\ref{SIII}. 
In Sec.~\ref{SIV}, we compare the ponderomotive wiggler potential with the classical diamagnetic potential associated with the $\mu \nabla B$ force and the centrifugal potential associated with the plasma rotation. The ordering of these various confining potentials and their respective advantages and drawbacks are discussed. The last section summarizes our new results.

\section{Interaction between a rotating magnetized plasma and an azimuthal wiggler}\label{SII}

In this section, we derive the leading order potential energy of a particle performing a cycloid motion with a weak azimuthal wiggler. We start with a Newtonian derivation of this average force using a one-dimensional model, illustrating the ponderomotive effect of the magnetostatic wiggler on a rotating particle. This derivation shows how the ponderomotive potential is independent of the rotation frequency, and that the Lorentz force generating the reflection is composed of the azimuthal velocity corresponding to average plasma rotation and the radial component of the wiggler. Later, we perform a Hamiltonian analysis, from which an average potential energy term, $\langle H_1\rangle$, naturally arises. The (average) force due to motion into such a potential is $\langle F_z\rangle=-d\langle H_1\rangle/dz$, as usual. In this subsection, particle motion is solved for all three dimensions. In order to achieve a similar azimuthal motion as in the Newtonian toy model, we use electromagnetic fields that are much stronger than the fields of the wiggler. In Hamiltonian language, the assumption is that the cycloid motion is generated by a Hamiltonian $H_0$ that is much larger than the Hamiltonian describing the interaction with the wiggler field, $H_1$.

\begin{figure}
	\centering
	\includegraphics[width =\columnwidth]{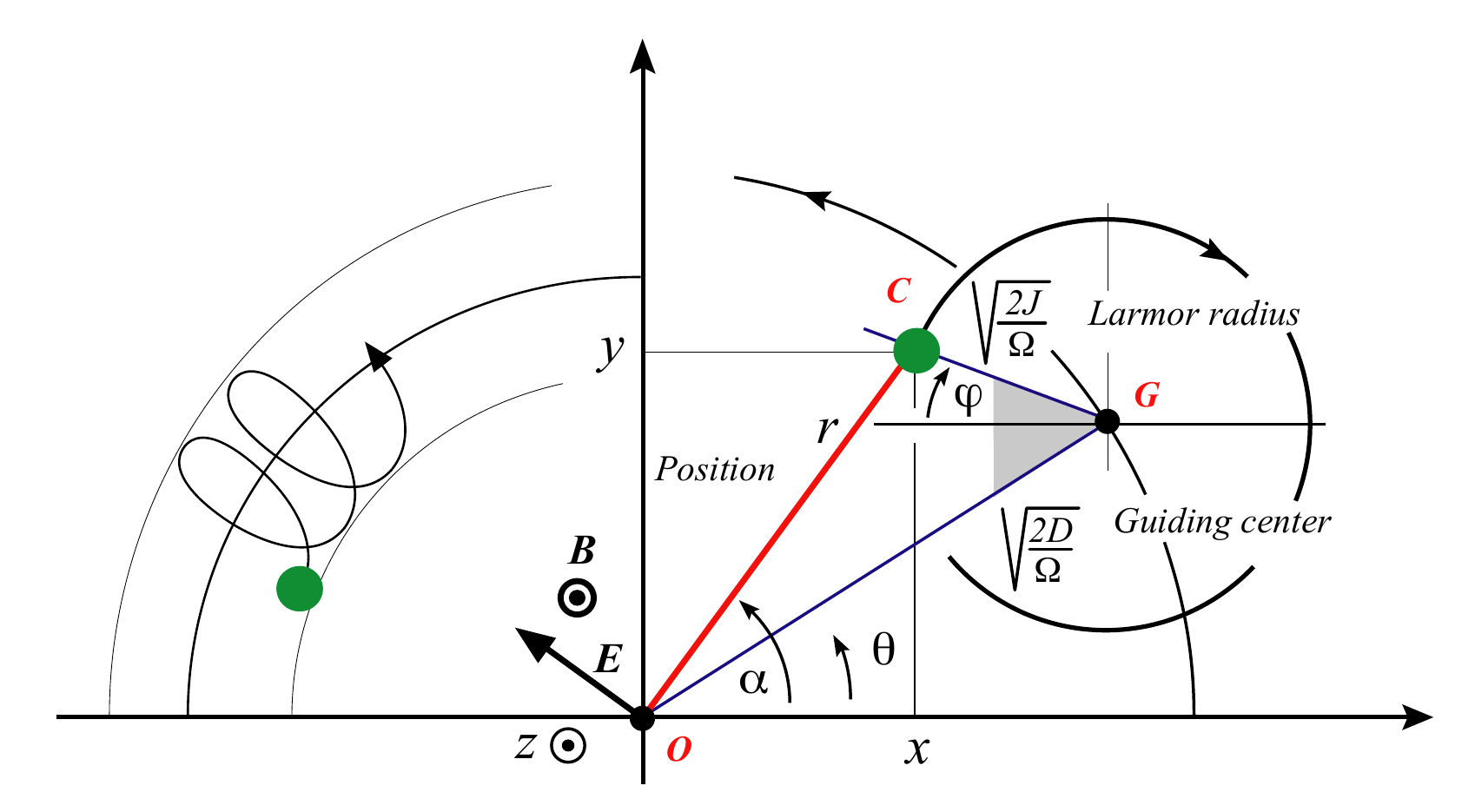}
	\caption{Physical meaning of the angle $\left( \varphi ,\theta \right) $ and actions $\left( J<D\right) $ variables in real $\left( x,y\right) $ space.
 Reproduced from  Rax, J. M., J. Robiche, R. Gueroult, and C. Ehrlacher. \textit{Kinetic Theory of Transport Driven Current in Centrally Fuelled Plasmas}, Physics of Plasmas \textbf{25}, 072503 (2018), with the permission of AIP Publishing.
}\label{fig:coords}
\end{figure}

To describe the fields configurations and the particles orbits, we use both a Cartesian set of coordinates $\left( x,y,z\right) $ and a polar one $\left(r,\alpha ,z\right) $ associated respectively with the basis $\left( \mathbf{e}_{x},\mathbf{e}_{y},\mathbf{e}_{z}\right) $ and $\left( \mathbf{e}_{r},\mathbf{e}_{\alpha },\mathbf{e}_{z}\right) $, such that $x=r\cos \alpha $ and $y=r\sin\alpha $. These sets of coordinates are illustrated in Fig. \ref{fig:coords} which displays a typical charged particle orbit in a rotating plasma.

The rotating plasma configuration is assumed to be a rigid body rotation of the Brillouin type. 
The Larmor radius is assumed to be smaller than the guiding center radius. 
In supersonic rotating mirrors, the plasma is usually confined in a thin cylindrical shell where the guiding center radius is larger than the Larmor radius, so the results obtained for this Brillouin configuration remain valid for a sheared rotation provided that the electric field and its radial derivative are adjusted to match locally the value of an equivalent Brillouin configuration. 
The potentials under which the particle is performing a cycloid motion are
    \begin{eqnarray}
        \mathbf{A} &=& \frac{1}{2}rB\mathbf{e}_\alpha = \frac{1}{2}(x\mathbf{e}_y-y\mathbf{e}_x)B, \\
        \Phi &=& \omega rA_\alpha = \frac{1}{2}r^2B\omega = \frac{1}{2}B\omega(x^2+y^2).\label{potential}
    \end{eqnarray}
The axial magnetic field $B$, and the $\mathbf{E}\times\mathbf{B}$ drift frequency is $\omega =\frac{1}{r} \frac{\mathbf{E}\times\mathbf{B}}{|B|^2}$ are constants.

The electric field is radial, and the magnetic field is axial,  as illustrated on Fig. \ref{fig:coords}. They can be expressed as 
    \begin{eqnarray}
        \mathbf{E} &=& - r \omega B \mathbf{e}_{r},  \label{chan1} \\
        \mathbf{B} &=&B\mathbf{e}_{z}.  \label{chan2}
    \end{eqnarray}
We define the following set of frequencies 
\begin{equation}
	\Omega _{c}=\frac{qB}{m}, \ \Omega_B =\sqrt{\Omega _{c}^{2}+4\omega \Omega_c}, \ \Omega _{\pm }=-\frac{1}{2}(\Omega _{c}\pm \Omega_B).  \label{omeg3}
\end{equation}
which are the cyclotron frequency $\Omega_c$, the Brillouin frequency $\Omega_B$, and $\Omega _{\pm }$ are the usual slow and fast Brillouin modes definitions\cite{brillouinTheoremLarmorIts1945,davidsonPhysicsNonneutralPlasmas1990}, associated with the dynamics of a particle with mass $m$ and charge $q$. We constrain the electric field such that $\Omega_B^2>0$, that is $4\omega>-\Omega_{c}$. This condition ensures particles with charge $q$ and mass $m$ are confined in this field configuration and not accelerated radially out.

This magnetic field can be generated by a current sheet
\begin{gather}
    \mathbf{j}_\mathrm{axial\ B} = \frac{B}{\mu_0}\delta(r-R)\mathbf{e}_\alpha,
\end{gather}
where $\mu_0$ is the permeability of free space, and $\delta$ is the Dirac distribution (Dirac delta). 

We add another magnetic field, $\tilde{\mathbf{B}}$, called ``wiggler", to the configuration described in Eqs. (\ref{chan1}, \ref{chan2}).
\begin{equation}
    \tilde{\mathbf{B}} = \begin{cases}
        \tilde{B}(z)\left(\frac{r}{R}\right)^{n-1}\left[\sin\left(n\alpha\right)\mathbf{e}_r+\cos\left(n\alpha\right)\mathbf{e}_\alpha\right],&\ r< R\\
        \tilde{B}(z)\left(\frac{R}{r}\right)^{n+1}\left[\sin\left(n\alpha\right)\mathbf{e}_r-\cos\left(n\alpha\right)\mathbf{e}_\alpha\right],&\ r> R
    \end{cases}\label{rf3}
\end{equation}
where $n\in\mathbb{N}$ is a positive integer, and $\tilde B$ is the field strength at $r=R$, which may be a function of z. A vector potential for this magnetic field is
\begin{equation}
    \mathbf{a} = \begin{cases} 
      -\tilde B(z)\frac{R}{n}\left(\frac{r}{R}\right) ^{n}\cos \left(n\alpha\right) \mathbf{e}_{z}, &\ r< R \\
      -\tilde B(z)\frac{R}{n}\left(\frac{R}{r}\right) ^{n}\cos \left(n\alpha\right) \mathbf{e}_{z}, &\ r> R
    \end{cases}\label{rf2}.
\end{equation}
As a \textit{curl} of a vector potential, this magnetic field is divergence-less, and as such - physical.

If $\tilde B$ is a function of z, this is not a vacuum field, supported by currents in the plasma which scale as $d\tilde{B}/dz$. In order to perform the averaging procedure later, we shall require a large length scale over which $\tilde{B}$ is ramped-up, so the deviation from a vacuum field is going to be small. 

In the limit of no z dependence of the wiggler field, it can be generated by a surface current density 
\begin{equation}
\mathbf{j}_\mathrm{wiggler}=-\frac{2\tilde B}{\mu_0}\cos \left(n\alpha\right)\delta \left( r-R\right)\mathbf{e}_{z}.\label{wigg curr}
\end{equation}

Practically, this current density can be realized by a set of axial (along $z$) wires arranged as an $n$ multipolar configuration (squirrel cage configuration or early \textit{Ioffe bar} configuration also proposed to rotate a plasma for the purpose of mass separation) around a cylinder of radius $R$\cite{raxRotationInstabilitiesIsotope2016}. It is to be noted that permanent magnet wiggler can also be considered\cite{halbachDesignPermanentMultipole1980}.

If there is a z dependence to $\tilde{B}$, a ``return" current must be added to Eq. (\ref{wigg curr}) so it remains divergence-less. One type of closure might be
\begin{multline}
    \mathbf{j}_\mathrm{wiggler}=\left(-\tilde B(z)\cos \left(n\alpha\right)\mathbf{e}_{z}+\frac{r}{n}\dv{\tilde{B}}{z}(z)\sin \left(n\alpha\right)\mathbf{e}_\alpha\right)\\\times\frac{2}{\mu_0}\delta \left( r-R\right).
\end{multline}

These are clearly ideal smooth currents. A realistic implementation would likely be comprised of a finite number of wires, coils, or permanent magnets. The effects reported in this work are not affected by small scale oscillations in the wiggler, which are removed by the averaging procedure we employ.

For the magnetic field in Eq. (\ref{chan2}) alone, any surface tangential to $\mathbf{e}_z$ is a magnetic surface. Specifically, coaxial circular cylinders centered at the origin are magnetic surfaces. The electrostatic potential in Eq. (\ref{potential}) is constant on these surfaces. Adding the wiggler field $\tilde{\mathbf{B}}$ to the magnetic field in Eq. (\ref{chan2}) alters the topology of magnetic surfaces. Now, $(r/R)^n\cos(n\alpha) = $ Const. are magnetic surfaces, and the electrostatic potential in Eq. (\ref{potential}) is no longer constant on magnetic surfaces. As a result, $\mathbf{E}\cdot (\mathbf{B}+\tilde{\mathbf{B}})=\mathbf{E}\cdot \tilde{\mathbf{B}}\neq0$. 

A major caveat in this field configuration is our assumption that the electric field, which is produced by a uniform charge distribution in the plasma, remains as in Eq. (\ref{chan1}), and is not affected by the presence of the wiggler. The charge distribution in the plasma may be rendered non-uniform due to reflection of ions, especially at larger radius, where the reflecting force is most effective. The electric field may also be modified by magnetohydrodynamical effects, by which the plasma might rearrange itself to minimize the total $|\mathbf{E}\cdot {\mathbf{B}}|$. Additionally, we might suppose that an electrostatic perturbed potential, if it existed, would be of the form, $\phi(r,\alpha,z) = f(r,z)\cos (n\alpha)$, with $f(r,z)$ some function of $r, \ z$, with a magnitude small enough such that its effect on rapidly rotating ions would be averaged out.

\subsection{Newtonian derivation}

Imagine a charged particle whose motion is constrained to lie on a cylinder with of radius $R_G$. Beside having no radial velocity, we constrain the azimuthal component of its velocity to be a constant $v_\alpha\neq 0$. This particle now interacts with the magnetic field described by Eq. (\ref{rf3}), and we take $R_G<R$. The equation of motion in the unconstrained z direction is
\begin{gather}
    m\ddot{z} = - q v_\alpha \tilde B(z) \left(\frac{R_G}{R}\right)^{n-1}\sin(n\alpha),
\end{gather}
with the argument of the sine satisfying 
\begin{gather}
    \tilde \omega = n\dot \alpha = \frac{n v_\alpha }{R_G}.
\end{gather}

Under these assumptions, the time dependent equation of motion becomes
\begin{gather}
    \ddot{z} = v_0 \tilde \Omega(z)\sin(\tilde \omega t + \alpha_0),\label{classical pondero}
\end{gather}
where we encapsulated the constant pre-factor as $v_0 = - \tilde \omega \frac{R_G}{n}   \left(\frac{R_G}{R}\right)^{n-1}$, $\tilde \Omega= q\tilde B/m$ is the cyclotron frequency associated with the magnetic field $\tilde B$, which is the strength of the wiggler field at $r=R$, and $\alpha_0$ is some initial angle.

We assume the oscillation frequency is much larger than the change in the envelope of the oscillations 
\begin{gather}
    \tilde \omega \gg \frac{\dot z}{\tilde \Omega}\dv{\tilde \Omega}{z},\label{adiabaticity newtonian}
\end{gather} 
and separate the motion into a slow $z_0$ and a fast oscillating part $z_1$, such that $z_1\ll z_0$. Taylor expanding Eq. (\ref{classical pondero}) yields
\begin{gather}
    \ddot{z}_0+\ddot{z}_1 \approx v_0 \left[\tilde \Omega(z_0)+z_1\dv{\tilde \Omega}{z}(z_0)\right]\sin(\tilde \omega t + \alpha_0).
\end{gather}
The leading order solution for the fast motion is
\begin{gather}
    {z}_1 \approx -\frac{v_0 \tilde \Omega(z_0)}{\tilde \omega^2}\sin(\tilde \omega t + \alpha_0).
\end{gather}
The slow motion becomes
\begin{gather}
    \ddot{z}_0\approx  -\frac{v_0^2}{4\tilde \omega^2}\dv{\tilde \Omega^2}{z}(z_0)\left(1-\cos(2\tilde \omega t + 2\alpha_0)\right).
\end{gather}

This equation is in the form of the traditional ponderomotive force equations.  Thus, averaging over the fast oscillations, and multiplying by $\langle\dot z_0\rangle$ yields
\begin{gather}
    \frac{1}{2}\dv{ \langle\dot z_0\rangle^2}{t} =  -\frac{v_0^2}{4\tilde \omega^2}\dv{\tilde \Omega^2}{t}(\langle z_0\rangle).
\end{gather}
Substituting the constant $v_0$ eliminates the $\tilde \omega$ frequency from this expression,
\begin{gather}
    \frac{1}{2}m \langle\dot z_0\rangle^2 +\frac{1}{4}m \tilde \Omega^2(\langle z_0\rangle)\frac{R^2}{n^2}   \left(\frac{R_G}{R}\right)^{2n}=\mathrm{Const}. \label{pond pot newt}
\end{gather}
The first term in this expression is the kinetic energy of the slow motion, while the second term is the ponderomotive potential we seek. Particle reflection would occur if both Eqs. (\ref{adiabaticity newtonian}) and (\ref{pond pot newt}) are satisfied. The length scale over which the wiggler field is ramped up is a free parameter, and can be selected independently.  

In the following subsection we derive the same potential using a Hamiltonian approach. In this more complete derivation, the radial and azimuthal constraints are approximately realized by the electromagnetic fields in Eqs. (\ref{chan1}, \ref{chan2}). 

\subsection{Hamiltonian analysis}
The unperturbed Hamiltonian $H_{0}$ of a rotating plasma in this configuration is the usual sum of the kinetic energy $m\mathbf{v}^{2}/2$ of the ion plus its potential energy $q\Phi \left( \mathbf{x}\right) $, 
\begin{equation}
H_{0}\left( \mathbf{p},\mathbf{x}\right) =\frac{1}{2}m\mathbf{v}^{2}+q\Phi =\frac{1}{2m}\left[ \mathbf{p-}q\mathbf{A}\left( \mathbf{x}\right) \right]^{2}+q\Phi \left( \mathbf{x}\right),  \label{h1}
\end{equation}
where $\mathbf{v}$ is the velocity and $\mathbf{p}=p_{x}\mathbf{e}_{x}+p_{y}%
\mathbf{e}_{y}+p_{z}\mathbf{e}_{z}$ is the canonical momentum conjugate to the position $\mathbf{x}$ = $x\mathbf{e}_{x}+y\mathbf{e}_{y}+z\mathbf{e}_{z}$ of the ion. This unperturbed Hamiltonian $H_{0}$ Eq. (\ref{h1}), associated with the electric and magnetic fields configuration described by Eqs. (\ref{chan1},\ref{chan2}), is expressed in
Cartesian coordinates as 
\begin{equation}
H_{0}=\frac{p_{x}^{2}+p_{y}^{2}+p_{z}^{2}}{2m} +\frac{\Omega _{c}}{2}%
\left( yp_{x}-xp_{y}\right) +\frac{\Omega_B ^{2}}{8}\left( x^{2}+y^{2}\right). \label{h2}
\end{equation}
This is a quadratic form of the Cartesian momentum and positions variables, thus $H_{0}$ is integrable. We consider the following change of variables adapted to the geometry of the problem, 
\begin{eqnarray}
x &=&\sqrt{\frac{2}{m\Omega_B}}\left(\sqrt{D}\cos \theta -\sqrt{J}\cos\varphi\right),\label{x}\\
y&=&\sqrt{\frac{2}{m\Omega_B}}\left(\sqrt{D}\sin \theta +\sqrt{J}\sin \varphi\right) ,\label{y}\\ 
p_{x} &=&\sqrt{\frac{1}{2}m\Omega_B}\left(-\sqrt{D}\sin \theta +\sqrt{J}\sin \varphi\right) ,\label{prxx1} \\
p_{y}&=&\sqrt{\frac{1}{2}m\Omega_B}\left(\sqrt{D}\cos \theta +\sqrt{J}\cos \varphi\right).\label{prxx2}
\end{eqnarray}
The actions variables $J$ and $D$ can be interpreted in terms of the guiding center radius $R_{G}$ and Larmor radius $\rho$ of the particle motion as 
\begin{equation}
R_{G}=\sqrt{\frac{2D}{m\Omega_B}}, \ \rho =\sqrt{\frac{2J}{m\Omega_B}},
\end{equation}
where $J\geq 0$, $D\geq 0$, and the angle variables $\theta\in\left[0,2\pi\right]$ and $\varphi\in\left[0,2\pi\right]$. This canonical transform was
already used in studies on transport driven currents\cite{raxKineticTheoryTransport2018}.

Through a simple substitution of Eqs. (\ref{x}, \ref{y}, \ref{prxx1}, \ref{prxx2}) in Eq. (\ref
{h2}) we express the Hamiltonian $H_{0}$ 
\begin{equation}
H_{0}=\frac{1}{2m}P^{2}+\Omega _{-}D-\Omega _{+}J,\label{h3}
\end{equation}
where we re-labeled $P=p_z$.
This expression is independent of the angles $\left( \varphi ,\theta,z\right) $ as expected. This particular canonical transform displays two advantages to set up the ion dynamics study (\textit{i}) in the dynamics generated by $H_0$ the actions are independent of time $d\left( J,D,P\right) /dt=0$, (\textit{ii}) the actions $\left( J,D,P\right) $ and angles $\left( \varphi ,\theta ,z\right) $ have a simple geometrical interpretation illustrated on Fig. \ref{fig:coords}.

The cyclotron (kinetic) part of the energy $-\Omega _{+}J \approx \frac{1}{2}m\Omega_c^2\rho^2$ is always positive although the drift (potential) part $\ \Omega _{-}D \approx \frac{1}{2}m\omega\Omega_cR_G^2$ can be either positive or negative, depending on the direction of the electric field and the sign of the particle electric charge. Hamilton's equations lead to the expected classical Brillouin results describing the uniform drift rotation of the guiding center around the axis of the configuration and the uniform cyclotron rotation around the that magnetic field line.
\begin{eqnarray}
    \frac{d\theta}{dt} =\Omega_-,\\
    \frac{d\varphi}{dt}=-\Omega_+.
\end{eqnarray}

The minus sign for the fast (cyclotron) rotation is due to the choice of a clockwise angle for $\varphi $ (the counterclockwise choice for $\theta $). We have thus identified a convenient set of canonical angles $\boldsymbol{\varphi} = \left( \varphi ,\theta ,z\right) $ and actions $\boldsymbol{J} = \left( J,D,P\right) $ variables describing the ion interaction with the electric and magnetic filed given by Eqs. (\ref{chan1}, \ref{chan2}).

The Hamiltonian $H$ describing the interaction of a particle with the DC confining fields, Eqs. (\ref{chan1}, \ref{chan2}), and the wiggler field, Eq. (\ref{rf3}), is given by 
\begin{eqnarray}
    H &=& H_0 + H_1,\\
    H_1 &=& -\frac{Pqa_z}{m}+ \frac{q^2a_z^2}{2m}\label{H_1}
\end{eqnarray}
Where $H_1$ is $H_1 =\left[\left( \mathbf{p}-q\mathbf{A}-q\mathbf{a}\right) ^{2}-\left( \mathbf{p}-q\mathbf{A}\right) ^{2}\right]/2m$ the kinetic energy term not already contained in $H_0$. 
In our case $\mathbf{A}\cdot \mathbf{a}=0$ and $\mathbf{p}\cdot \mathbf{a} = p_za_z$. 

Looking at particle motion confined within $r<R$ in its entirety,
\begin{eqnarray}
-\frac{Pqa_z}{m} &=&P\tilde \Omega \frac{R}{n}\left(\frac{r}{R}\right) ^{n}\cos \left(n\alpha\right)\label{dipolar},  \\
\frac{q^2a_z^{2}}{2m} &=&\frac{1}{4} m\tilde \Omega^2\frac{R^2}{n^2}\left(\frac{r}{R}\right) ^{2n}\left(1+\cos\left(2n\alpha\right)\right)\label{pond}.
\end{eqnarray}

The procedure employed here is to substitute the change of variables $(r, \ \alpha)\rightarrow(x,\ y) \rightarrow (\theta,\ \varphi,\ D,\ J)$ using Eqs. (\ref{x}) and (\ref{y}) into Eqs. (\ref{dipolar}) and (\ref{pond}). Because $(z,\ P)$ are unaffected by this change of variables, any z-dependence remains unchanged. 

In order to identify secular effects of the ponderomotive type we average over the unperturbed cyclotron and drift rotations ($\theta \sim \Omega _{-}t$ , $\varphi \sim \Omega _{+}t$). This averaging will give $\left\langle \mathbf{p}\cdot \mathbf{a}\right\rangle _{\theta ,\varphi }=0$
, and $\left\langle a^{2}\right\rangle _{\theta ,\varphi }\neq 0$. The dipolar coupling described by $\mathbf{p}\cdot \mathbf{a}$ can give a resonant ponderomotive effect if we set up a second order perturbative expansion, but this term is always far smaller than the $\left\langle a^{2}/2\right\rangle _{\theta ,\varphi }$ term except very near resonances between the cyclotron and drift motion, where the averaging methods becomes questionable\cite{lichtenbergRegularStochasticMotion1983a}. We will study this minor contribution to the dynamics in a forthcoming paper and evaluate its very specific properties associated with mass selection and non reciprocity. The dominant effect of the wiggler is thus $\left\langle q^2a^{2}/2m\right\rangle _{\theta ,\varphi }$ and, as shown below and in the next sections, it sets up an axial ponderomotive potential providing end plugging in a rotating plasma column.

In order to perform the averaging of the perturbation $H_1$ over the motion generated by $H_0$, we assume $H_1/H_0\sim \epsilon<1$. To leading order, the change in energy
\begin{equation}
    \frac{dH_1}{dt}= \pdv{H_1}{\boldsymbol{J}}\cdot\dv{\boldsymbol{J}}{t}+\pdv{H_1}{\boldsymbol{\varphi}}\cdot\dv{\boldsymbol{\varphi}}{t},
\end{equation}
is due to the angular motion, as the evolution of the actions is slow, $d\boldsymbol{J}/dt = -\partial H / \partial \boldsymbol{\varphi}=  -\partial H_1 / \partial \boldsymbol{\varphi}\sim\mathcal{O}(\epsilon)$, whereas the evolution of the angles is fast $d\boldsymbol{\varphi}/dt = \partial H/\partial \boldsymbol{J} \approx\partial H_0/\partial \boldsymbol{J}\sim\mathcal{O}(1)$. We want to average over the angles $\theta$ and $\varphi$, so we require time scale separation such that
\begin{eqnarray}
    \left|\pdv{H_1}{z}\frac{P}{m}\right|&\ll&\left|\pdv{H_1}{\theta}\Omega_-\right|,\label{ad con1}\\
    \left|\pdv{H_1}{z}\frac{P}{m}\right|&\ll&\left|\pdv{H_1}{\phi}\Omega_+\right|. \label{ad con2}
\end{eqnarray}

Neglecting fringing fields, we take the wiggler field to be a function of $z$, $\tilde \Omega (z)$. The adiabaticity conditions Eqs. (\ref{ad con1}, \ref{ad con2}) are then
\begin{equation}
    \left|\frac{P}{m}\dv{\ln \tilde \Omega}{z}\right|\ll\left|n \Omega_\pm\right|.
\end{equation}
Equivalently, using $L_z^{-1} =| d\ln \tilde \Omega/dz|$ as the ramp-up length scale, which can be a function of z, the more stringent condition in the common limit of $\omega\ll \Omega_c$, is 
\begin{gather}
    v_z\ll n \Omega_- L_z\approx n \omega L_z.\label{adiabaticity}
\end{gather}

To leading order, the cosine terms in Eqs. (\ref{dipolar}, \ref{pond}) and in the perturbed potential $\phi$ average to zero. This is evident in the small gyroradius limit $R_G\gg\rho$, where $\alpha \approx \theta\sim \Omega_-t$, but this is also true in the general case.

We can use Eqs. (\ref{x}, \ref{y}) to evaluate the reminder of Eq. (\ref{pond}) by
\begin{equation}
    r^{2}=\frac{2}{m\Omega_B}\left[\left(D+J\right)-2\sqrt{DJ}\cos \left( \theta+\varphi \right)\right]. \label{tri}
\end{equation}
Employing the binomial theorem,
\begin{multline}
    r^{2n}=\left(\frac{2}{m\Omega_B}\right)^n\sum_{\ell=0}^{n}(-1)^{\ell}\mathcal{C}_{\ell}^{n}\left(D+J\right)^{n-\ell}\left(DJ\right)^{\ell/2}\\
    \times\left[2\cos\left(\theta+\varphi \right)\right]^\ell,
\end{multline}
where $\mathcal{C}_{\ell}^{n}= \binom{n}{\ell}=n!/\ell!\left( n-\ell\right)!$ are the binomial coefficients.
Expanding the cosine and keeping only the terms that do not depend on the angles
\begin{equation}
    \langle r^{2n}\rangle=\left(\frac{2}{m\Omega_B}\right)^n\sum_{\ell=0,2,4,...}^{n}\mathcal{C}_{\ell}^{n}\mathcal{C}_{\ell/2}^{\ell}\left(D+J\right)^{n-\ell}\left(DJ\right)^{\ell/2}.
\end{equation}

The Hamiltonian describing the leading order effect of the wiggler on the particle motion is $H$ = $H_{0}+\langle H_1\rangle $ where 
\begin{equation}
\langle H_1\rangle =\frac{1}{4}m\tilde\Omega^{2} \frac{R^2}{n^2}\left( 2\frac{D+J}{m\Omega_B R^{2}}\right)^{n}\sum_{\ell=0,2...}^{n}\mathcal{C}_{\ell}^{n}\mathcal{C}_{\ell/2}^{\ell}\left(\frac{\sqrt{DJ}}{D+J}\right) ^{\ell}.\label{H09}
\end{equation}

Particles entering the wiggler region will be reflected if their axial energy outside the wiggler, $ m v_z^2/2 = P^2/2m <\langle H_1\rangle$, and the wiggler ramp up is adiabatic, as described in Eq. (\ref{adiabaticity}).

Hamilton's equation along the $z$ direction is then given by 
\begin{equation}
\frac{dP}{dt}=-\frac{\partial H}{\partial z}=2\dv{\ln\tilde \Omega}{z}\left\langle H_1\right\rangle \label{ppp}.
\end{equation}
This relation describes an attractive or repulsive force along $z$, according to the sign of $d\tilde \Omega/dz$. This analytical results is checked numerically in the next section. 

\section{Numerical study}\label{SIII}
In order to validate the prediction for the average behavior of the particle as it enters the wiggler region, we use a Boris pusher\cite{boris1970relativistic,stoltzEfficiencyBorislikeIntegration2002,qinWhyRelaxBorisAlgorithm2013} implemented in the LOOPP code used in earlier studies of time-dependent ponderomotive forces\cite{ochsNonresonantDiffusionAlpha2021a,ochsPonderomotiveRecoilElectromagnetic2023} to time-step a particle through the Lorentz force
\begin{gather}
    m \ddot{\mathbf{x}} = q(\mathbf{E} + \dot{\mathbf{x}}\times \mathbf{B}),
\end{gather}
with the electromagnetic fields given in cylindrical coordinates as
\begin{gather}
        \mathbf{E} = - r \omega B \mathbf{e}_{r}, 
\end{gather}
\begin{multline}
        \mathbf{B} =B\mathbf{e}_{z}  \\ +\frac{\tilde{B_0}}{1+\exp(-z/L_z)}\left(\frac{r}{R}\right)^{n-1}\left[\sin\left(n\alpha\right)\mathbf{e}_r+\cos\left(n\alpha\right)\mathbf{e}_\alpha\right].\label{numB} 
\end{multline}
With $B,\ \omega,\ \tilde B_0,\ L_z, R,\ n$ constants given in Table \ref{tab:parameters} for the four example trajectories, and $m,\ q$ being the deutron mass and elementary charge, respectively. 

The second line in (\ref{numB}) is the magnetic field of an external wiggler. The z dependence of the field satisfies $\forall z\in \mathbb{R}:d\ln \left(\tilde B_0 \cdot (1+\exp(-z/L_z))^{-1})\right)/dz\le L_z^{-1}$. The parameter $L_z$ sets the ramp-up length scale in Eq. (\ref{adiabaticity}). 

In the limits outlined above, the particle decelerates as it enters the wiggler region. It suffices to satisfy the reflection conditions for using the particle axial velocity value outside of the wiggler. The reflection conditions are: adiabatic dynamics (\ref{adiabaticity num}), and parallel energy being lower than the potential (\ref{energy}):
\begin{eqnarray}
    v_{z0} &\ll& n \omega L_z,\label{adiabaticity num}\\
    \frac{1}{2}mv_{z0}^2 &<& \langle H_1 \rangle .\label{energy}
\end{eqnarray}

\begin{table}
    \centering
    \begin{tabular}{|c|c|c|c|c|}
        \hline
        Parameter& Case I & Case II& Case III& Case IV\\ \hline
        $B$ &\multicolumn{4}{c|}{$10[T]$}\\ \hline
        $\omega$ & \multicolumn{2}{c|}{$28[M rad/s]$}&\multicolumn{2}{c|}{$280[k rad/s]$}\\ \hline 
        $\tilde{B}_{0}$& \multicolumn{4}{c|}{$1[T]$}  \\ \hline
        $n$& \multicolumn{2}{c|}{$2$}&\multicolumn{2}{c|}{$4$} \\\hline
        $R$&\multicolumn{4}{c|}{$1[m]$}\\\hline
        $L_z$ & \multicolumn{2}{c|}{$0.1 [m]$}&$3 [m]$&$1[m]$\\ \hline
        $R_{G}$&\multicolumn{2}{c|}{$ 0.3[m]$} &\multicolumn{2}{c|}{$0.65[m]$}\\ \hline
        $\rho$&$5[cm]$& \multicolumn{3}{c|}{$ 0[m]$}  \\ \hline
        $v_{z0}$& \multicolumn{4}{c|}{$1.38[Mm/s]$}\\ \hline
        $\Omega_-D$& \multicolumn{2}{c|}{$13.3 [MeV]$}& \multicolumn{2}{c|}{$591.8[keV]$}\\\hline
        $\Omega_+J$& $7.0[MeV]$&\multicolumn{3}{c|}{$0[eV]$}\\\hline
        $\frac{1}{2}m v_{z0}^2$&\multicolumn{4}{c|}{$20[keV]$}\\ \hline
        $\frac{1}{2}m v_{z0}^2 / \langle H_1 \rangle$& $0.74$& $0.82$&\multicolumn{2}{c|}{$0.84$}\\ \hline
        $v_{z0} /n L_z \Omega_-$& \multicolumn{2}{c|}{$0.26$}&$0.41$&$1.23$ \\ \hline
        particle & \multicolumn{4}{c|}{deutron} \\\hline        
    \end{tabular}
    \caption{Parameters for single particle simulations.}
    \label{tab:parameters}
\end{table}

\begin{figure*}
    \centering
    \includegraphics[width = \textwidth]{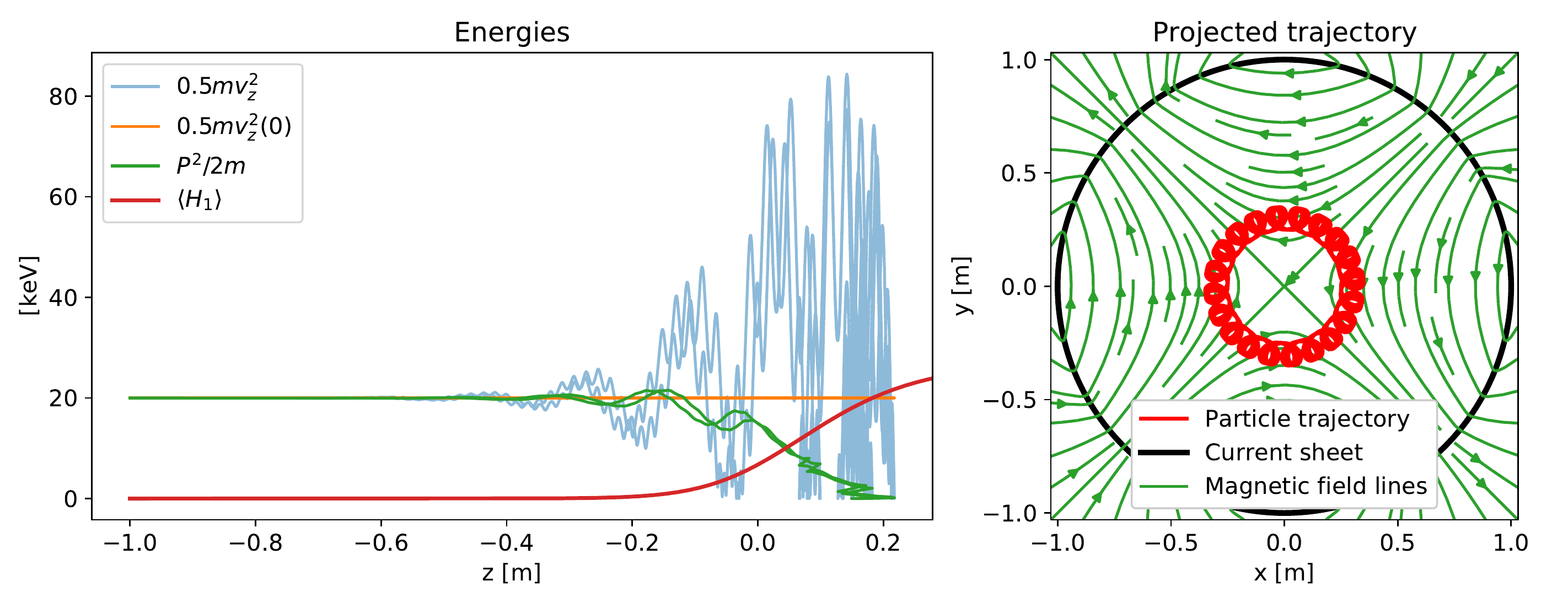}
    \caption{Energies as a function of axial position and trajectory in the $x-y$ plane, case I (parameters in Table \ref{tab:parameters}). Left: In blue - energy in the axial degree of freedom. In orange - the initial energy in the axial direction. In green - the energy in the axial canonical momentum. In red - the analytic expression for the potential barrier. Right: In red - the projection of the particle trajectory on the $x-y$ plane. In black - the device radius, where the current sheets generating the wiggler field are located. In green - the wiggler magnetic field lines. Reflection occurs when the energy in the canonical momentum is zeroed out.}    
    \label{fig:trajectory1}
    \includegraphics[width = \textwidth]{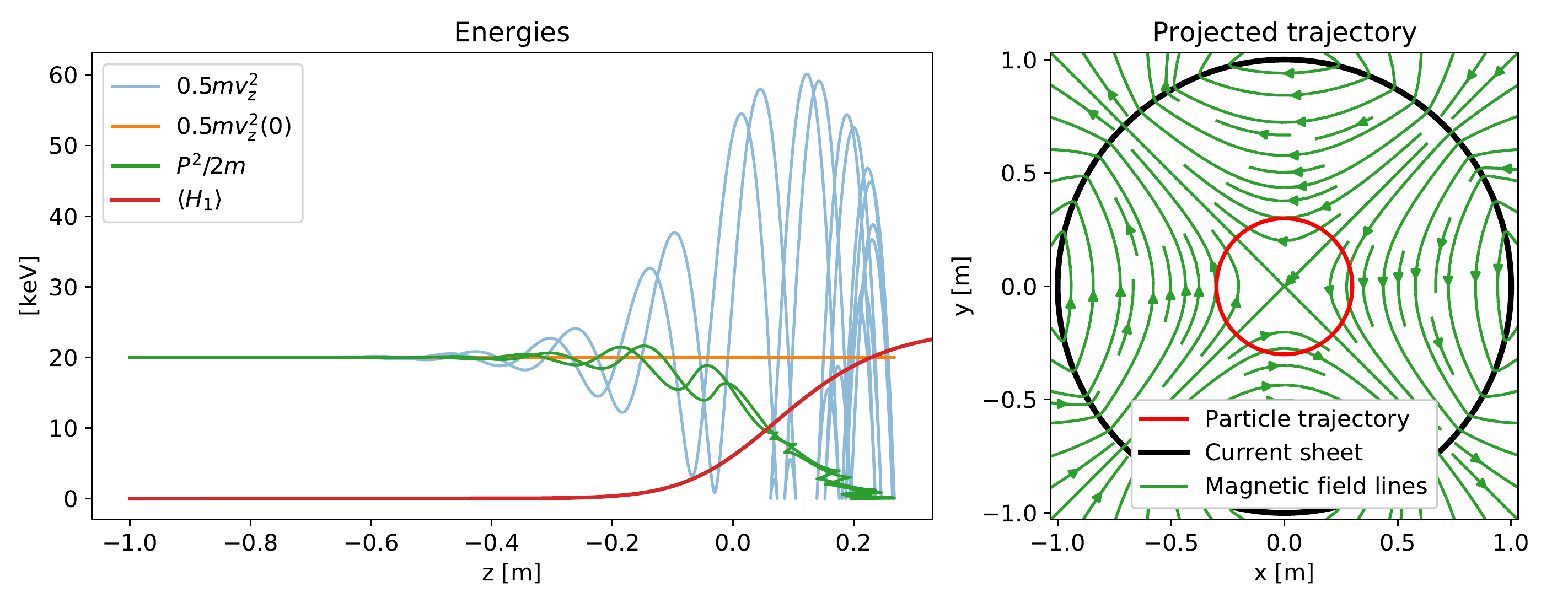}
    \caption{Energies as a function of axial position and trajectory in the $x-y$ plane, case II (parameters in Table \ref{tab:parameters}). Compared to Fig. \ref{fig:trajectory1}, no energy is in the $J$ degree of freedom, all the energy in the perpendicular motion is in the $D$ degree of freedom. For this configuration, the $\langle H_1 \rangle >20[keV]$ for more than $91\%$ of the cylinder cross-sectional area.}
    \label{fig:trajectory2}
\end{figure*}

The simulations for cases I and II are presented in Figures \ref{fig:trajectory1} and \ref{fig:trajectory2}. In these cases, two deutrons of $20[keV]$ axial energy are reflected from the same electromagnetic field configuration. The rotation frequency, $\omega$, wiggler periodicity, $n$, and axial ramp up length scale $L_z$ are chosen such that the motion is adiabatic, with $v_{z0} /n L_z \Omega_-=0.26\ll1$. The deutron in case I has a nonzero Larmor radius, $\rho = 5[cm]$, and performs a cycloid motion around the axis of the configuration. The ratio of the fast and slow Brillouin frequencies is not an integer, and the projection of the trajectory on the $x-y$ plane does not trace the same exact path. Case II is a cleaner picture of the same initial conditions, with $\rho =0[m]$. In both cases, the reflected particle has the same axial energy as it had before reflection. Case I and II show that particles of $20 [keV]$ or less whose initial conditions are in the $\approx 90\%$ area of the cylinder would be reflected.

\begin{figure*}
    \centering
    \includegraphics[width = \textwidth]{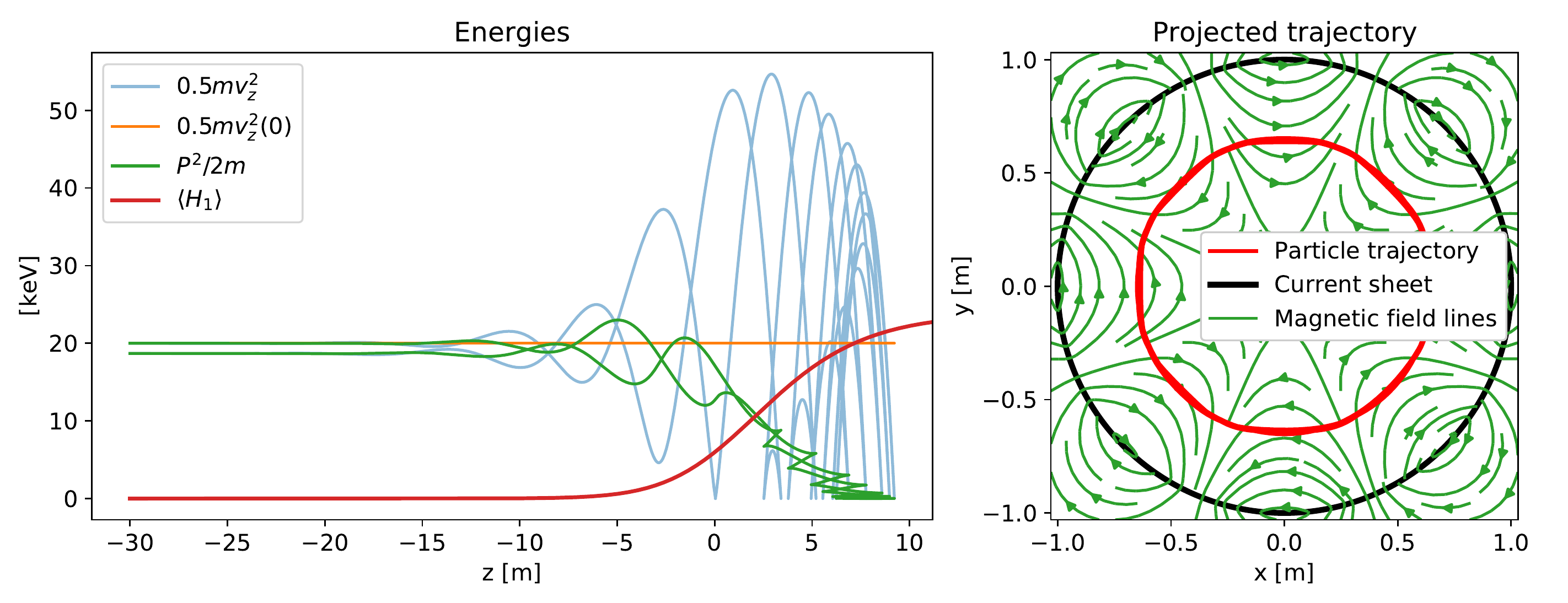}
    \caption{Energies as a function of axial position and trajectory in the $x-y$ plane, case III (parameters in Table \ref{tab:parameters}). Compared to Fig. \ref{fig:trajectory2}, this motion is almost non-adiabatic.}    
    \label{fig:trajectory3}
    \includegraphics[width = \textwidth]{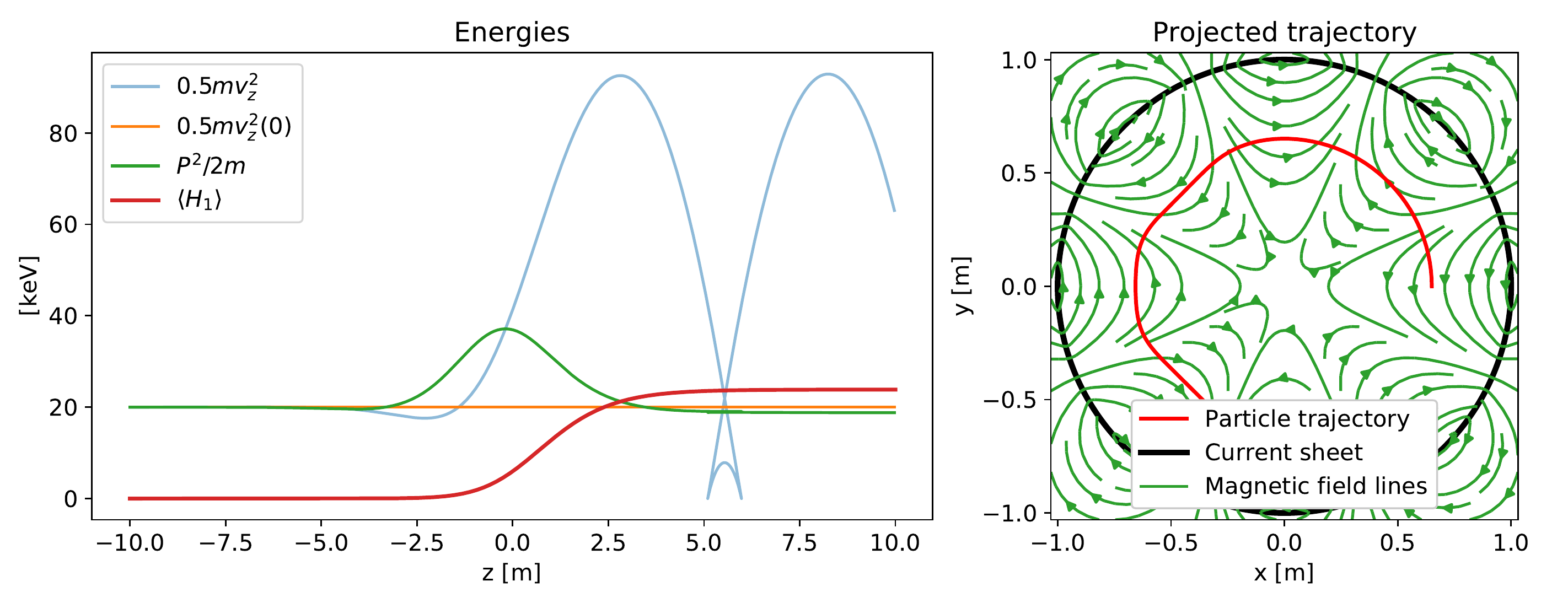}
    \caption{Energies as a function of axial position and trajectory in the $x-y$ plane, case IV (parameters in Table \ref{tab:parameters}). Compared to Fig. \ref{fig:trajectory2}, the particle interaction with the wiggler field is non-adiabatic.}
    \label{fig:trajectory4}
\end{figure*}

Cases III and IV, presented in Figures \ref{fig:trajectory3} and \ref{fig:trajectory4}, are examples of adiabaticity near-breaking and breaking. In these cases, the plasma rotation frequency is reduced to $280[k rad/s]$, which is more easily achievable in practice. In order to use a small $L_z$, which is $3[m]$ in case III, the wiggler periodicity is increased to $4$. This results in $\approx 57\%$ of the cylinder area to have a sufficient potential height to reflect $20[keV]$ deutrons. In the case of this not completely adiabatic interaction $v_{z0} /n L_z \Omega_-=0.41<1$, the axial energy of the reflected particle is somewhat different than the origianl.

If we use a smaller $L_z$, some initial angles would lead to particles ``tunneling" through the potential barrier, for example $L_z=1[m]$ in case IV. 

Reflection occurs when the energy in the axial canonical momentum, $P^2/2m$ reaches zero. The deviation of the reflection point from the expected curve $\langle H_1 \rangle$ is as large as the amplitude of oscillation of the energy in the axial canonical momentum, which $\langle H_1 \rangle$ averages out. The axial instantaneous energy $0.5 m v_z^2$ has a large spread around its initial value due to interaction with the wiggler field. The instantaneous velocity of the particle becomes negative briefly several times well before particle reflection.

\section{Weak electric field and small gyro radius limit}
If we consider the weak electric field approximation, $\omega<\Omega _{c}$, and the small Larmor radius approximation, $\rho<R_{G}$. Within the framework of these ordering we can expand $\Omega _{\pm }$ to express 
\begin{eqnarray}
\Omega _{-}D &\approx &\frac{1}{2}m\Omega_c\omega R_G^2+\frac{1}{2}m\omega^2R_G^2 
, \\
-\Omega _{+}J &\approx &\frac{1}{2}m\Omega_c^2 \rho^2+\frac{3}{2}m\Omega_c \omega \rho^2-\frac{1}{2}m\omega^2 \rho^2,
\end{eqnarray}
where we have neglected higher orders inertial terms. Expressing $\omega = -\frac{1}{B}\dv{E_r}{r}$, the energy 
\begin{multline}
    H_{0}=\frac{P^2}{2m}-\Omega _{+}J+\Omega _{-}D\\\approx \frac{1}{2}mv_\parallel^2+\frac{1}{2}mv_c^2+\frac{1}{2}mv_{\mathbf{E}\times\mathbf{B}}^2+q\Phi
\end{multline}
appears as a decomposition of the parallel energy, the energy of the cyclotron motion and the $\mathbf{E}\times\mathbf{B}$ drift energy, plus the electric potential energy $q\Phi(R_G)$. Notably, the energy do not depend on the angle $\theta+\varphi$, due to the chosen electric field profile ($E_r = r \cdot \mathrm{Const.}$), so no averaging is needed in order to separate the perpendicular energy into the drift and cyclotron parts.

\begin{figure}
    \centering
    \includegraphics[width=\columnwidth]{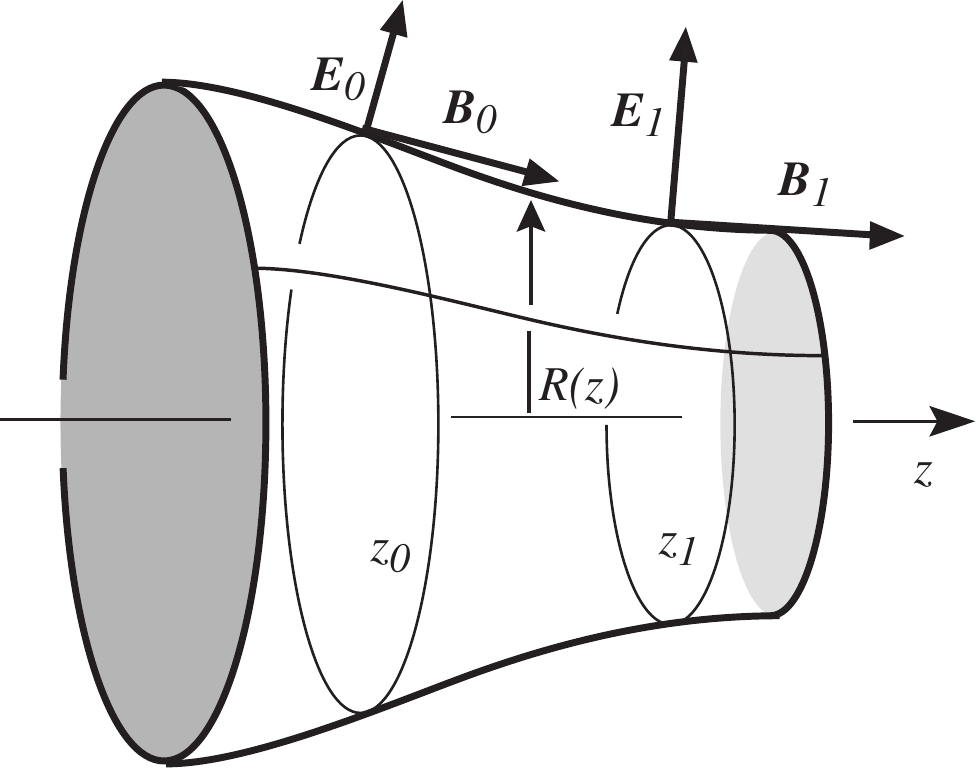}
    \caption{A rotating axisymmetric plasma
column with two section $z_{0}$ and $z_{1}$ along a field line.}
    \label{fig:axisym2}
\end{figure}

\begin{figure}
    \centering
    \includegraphics[width = \columnwidth]{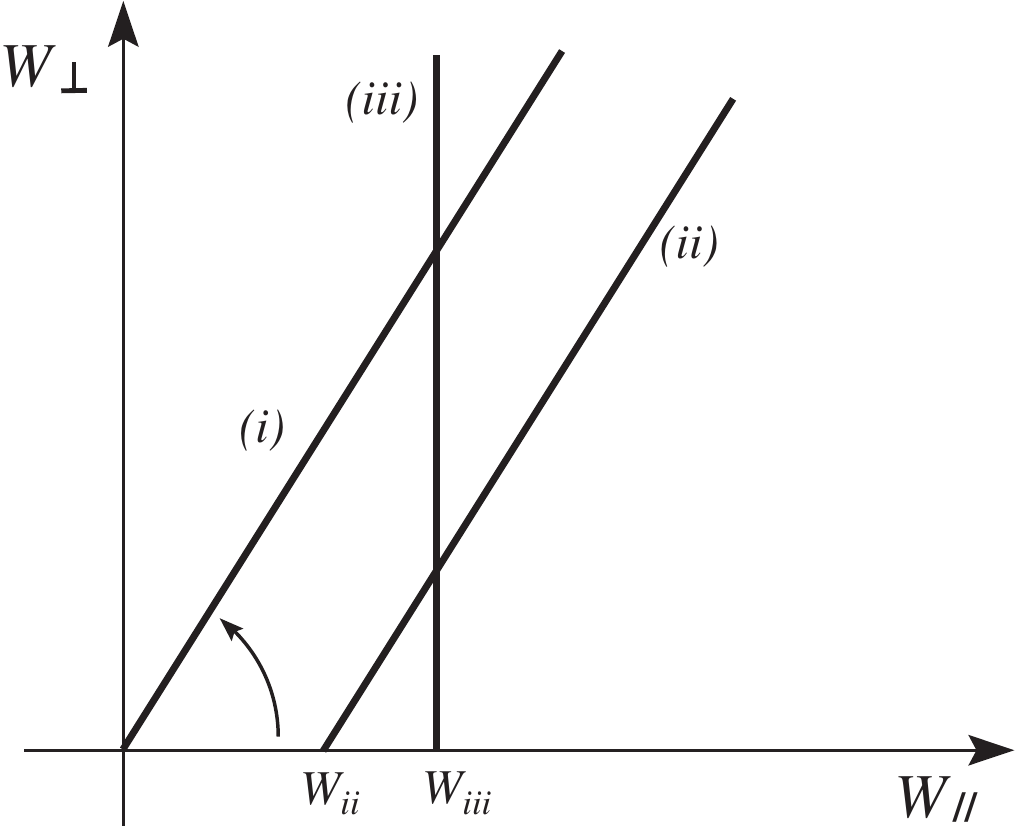}
    \caption{Boundaries of confined regions in the parallel
and cyclotron energy space at $z_{0}$, (\textit{i}) classical mirrors, (\textit{ii}) rotating mirrors and (\textit{iii}) end plugging with an azymuthal wiggler. $\cot \Theta =\mathcal{R}-1$, $W_{ii}=\frac{1}{2}mE_{0}^{2}/B_{0}^{2}\left(1-\mathcal{R}^{-1}\right)$.}
    \label{fig:trap}
\end{figure}

The charged particle dynamics associated with $H_{0}$ is more easily analyzed in a rotating frame where the plasma is \textit{at rest}, that is, rotating with $\omega$ frequency. In this frame the electric field cancels but we have to take into account the centrifugal potential and the Coriolis gyroscopic coupling. The Coriolis coupling appears as an additional effective magnetic field which can be neglected compared to $B$ in the weak electric field approximation. Thus, if we introduce the magnetic moment $\mu =mv_{c}^{2}/2B$
, the Hamiltonian $H_{0}$ in the rotating frame becomes $H_{0}^{\prime }$ which the sum of a kinetic term $mv_{\parallel }^{2}/2$ plus the diamagnetic and centrifugal potentials. 
\begin{equation}
H_{0}^{^{\prime }}=\frac{1}{2}mv_{\parallel }^{2}+\mu B-\frac{1}{2}m\omega^{2}R_{G}^{2}\text{.}  \label{ham567}
\end{equation}
Now let us consider an axisymmetric magnetic field, illustrated on Fig. \ref{fig:axisym2}, where we neglect the ambipolar potential along the field lines. 
The magnetic surfaces are described by the relations $r=$ $R\left( z\right)$. We define the mirror ratio between two sections $z=z_{0}$ and $z=z_{1}$ as $\mathcal{R}=B_1/B_0$ where $B_{0}=B\left( z_{0}\right) $, $B_{1}=B\left( z_{1}\right) $. Between these two sections $z=z_{0}$ and $z=z_{1}$, the conservation of the magnetic flux and the isorotation law can be expressed as 
\[
B_{0}R_{0}^{2}=B_{1}R_{1}^{2}\text{ , }\frac{E_{0}}{B_{0}R_{0}}=\frac{E_{1}}{%
B_{1}R_{1}}\text{.}
\]
The conservation of the energy $H_{0}^{\prime }$ Eq. (\ref{ham567}) within the adiabatic approximation, leads to the change of parallel velocity
between the sections $z=z_{0}$ and $z=z_{1}$ 
\begin{equation}
v_{\parallel 0}^{2}-v_{\parallel 1}^{2}=v_{c0}^{2}\left( \mathcal{R}%
-1\right) +\frac{E_{0}^{2}}{B_{0}^{2}}\left( 1-\frac{1}{\mathcal{R}}\right) 
\text{.}  \label{rotf23}
\end{equation}
For a straight plasma column with radius $R$ interacting with an
adiabatically tapered (external) wiggler with amplitude $\tilde B_2(z) = \tilde B\left( z\right) $ the
change of parallel velocity between two sections $z=z_{0}$ and $z=z_{1}$ is
given 
\begin{equation}
v_{\parallel 0}^{2}-v_{\parallel 1}^{2}=\frac{q^{2}}{2m^{2}}\left( \frac{R_{G}}{R}\right) ^{2n}\left( \frac{R}{n}\right) ^{2}\left[ \tilde{B}^{2}\left(z_{0}\right) -\tilde{B}^{2}\left( z_{1}\right) \right]   \label{www2}
\end{equation}
The relations Eqs. (\ref{rotf23}, \ref{www2}) allows two compare the
confinement properties of (\textit{i}) classical mirrors, (\textit{ii}) rotating mirrors and (\textit{iii}) end plugging with an external azimuthal wiggler. This comparison is clearly illustrated on Fig. \ref{fig:trap}. 
where we draw the boundary between confined and unconfined particles through the reflection requirement at $z_{1}$: $v_{\parallel 1}^{2}=0$. In this figure $W_{iii}$ is given by by the lower of $m/2$ times the right hand side of Eqs. (\ref{www2}), and $ 0.05\frac{1}{2}m  n^2 L_z^2 \omega^2$.

\section{Comparison of the ponderomotive potential with the diamagnetic and
centrifugal ones}\label{SIV}

If we consider a rotating mirror configuration there are two methods to mitigate the end losses : (\textit{i}) magnetic mirror point along the axis, or (\textit{ii}) an azimuthal wiggler around the axis. In this section we will compare and discuss the three potential associated respectively with the centrifugal force and these two end plugging schemes. The Hamiltonian of the problem clearly display these three energies
\begin{equation}
H-\frac{P^2}{2m}=-\Omega _{+}J+\Omega _{-}D+\left\langle H_1\right\rangle .
\end{equation}
The first term on the right hand side is (\textit{i}) the diamagnetic energy, the second term (\textit{ii}) the centrifugal energy and the third one (\textit{iii}) the leading order ponderomotive energy. Under the hypothesis $J\ll D$ the azimuthal wiggler ponderomotive potential is given by
\begin{equation}
\langle H_1\rangle = \left\langle \frac{q^{2}a^{2}}{2m}\right\rangle \approx \frac{q^{2}}{4m}\left( \frac{R_{G}}{R}\right) ^{2n}\left( \frac{R}{n}\right)^{2}\tilde{B}^{2}\left( z\right)   \label{www}
\end{equation}

For a classical magnetic mirror the $\mu B$ (dia)magnetic 
energy $-\Omega _{+}J=\left\langle \mu B\right\rangle $ is given by
\begin{equation}
\left\langle \mu B\right\rangle \approx\frac{1}{2}\frac{q^{2}}{m}\rho^{2}B^{2}\left( z\right) . \label{mug}
\end{equation}

The obvious advantage of wiggler end-plugging Eq. (\ref{www}) is the fact that it acts on the particles independently of their pitch angle although Eq. (\ref{mug}) displays the usual drawback associated with the loss cone.

The ratio of these two energies is
\begin{gather}
\frac{\left\langle H_1\right\rangle }{\left\langle \mu
B\right\rangle }\approx\frac{1}{2}\left( \frac{R_{G}}{R_2}\right) ^{2n}\left( \frac{R}{n\rho}\right) ^{2}\left( \frac{\tilde B}{B}\right) ^{2}
\end{gather}
and can be of the order one if $R_1\tilde{B}_1/\rho B\sim n$ and $R_{G}\sim R_1$ or $R_2\tilde{B}_2/\rho B\sim n$ and $R_{G}\sim R_2$. As $R_1+\rho \leq R_{G}\leq R_2-\rho$ bringing $R_1$ and $R_2$ close to one another increases this ratio favourably.

Finally the averaged potential energy $\left\langle q^{2}a^{2}/2m\right\rangle $ is to be compared with the centrifugal energy $\Omega _{-}D= m\Omega_{-}\Omega_B R_{G}^{2}/2$
\begin{gather}
    \frac{\left\langle H_1\right\rangle }{m\Omega _{-}\Omega_B R_{G}^{2}/2}\approx
    \frac{\tilde{\Omega}^{2}}{2\omega\Omega_c}\left( \frac{R_{G}}{R}\right) ^{2n}\left( \frac{R}{nR_G}\right)^{2}
\end{gather}

We can also compare the potentials produced by increasing $|B|$ to the same value. The added potential for $|B|=|B+\Delta B_\parallel|$ is $\mu\Delta B_\parallel$. Increasing the magnetic field to the same strength for a gyrocenter position $R_G$,
\begin{equation}
|B| = \sqrt{B^2 + \tilde B^2 \left(\frac{R_G}{R}\right)^{2n-2}} =B+\Delta B_\parallel 
\end{equation}

\begin{equation}
\frac{\left\langle H_1\right\rangle }{\mu \Delta B_\parallel}\approx \left( \frac{R_G}{n\rho}\right)^{2}\left(1+\frac{\Delta B_\parallel}{2B}\right)
\end{equation}
where $\rho$ is the Larmor radius before the interaction with the increased magnetic field $B+\Delta B_\parallel$.

Each drift surface would experience a different increase in $|B|$, owing to the radial depencence of the wiggler field.

\section{Conclusion}\label{SVI}

We offer a  magnetostatic end-plugging concept, using an azimuthal wiggler field, added to a strong axial field, such that the energy of the parallel motion is converted mostly to axial oscillations, resulting in reflection. 
The ponderomotive reflection is made possible because of the plasma rotation through static magnetic perturbations, as opposed to conventional ponderomotive barriers which involve reflection by imposed RF fields oscillating in time. The main assumption here is that surfaces of constant potential are not affected by the perturbation to the magnetic field. The plasma rotation provides for an oscillating field seen in the rotating particle reference frame.

This new type of reflection can be compared to magnetic mirror reflection, in which the parallel energy is converted into perpendicular motion. Magnetic mirrors fail to reflect charged particles with small perpendicular speeds; in contrast, the magnetostatic wiggler reflects such particles too.

As an adiabatic ponderomotive effect, it relies on slow changes in the envelope of the wiggler. This is a limitation of all effects of the ponderomotive type. Reflecting high energy particles, besides requiring a high ponderomotive potential barrier, would require construction of long devices with large rotation speeds.

Of course, any small transverse magnetic field can reflect particles by arcing them over half a gyro-radius. However, for weak transverse fields, that gyro-radius becomes large. 
In contrast, in the rotating plasma configuration, charged particles are reflected over transverse distances not comparable to the gyroradii in the weak wiggler field, but rather the comparable to the gyroradii in the axial guide field.  

Thus, the ponderomotive barrier offered here offers the advantages of: (\textit{i}) reflection of particles without relying on large perpendicular energies; (\textit{ii}) reflection over small transverse distances; and, most importantly, (\textit{iii}) reflection without the need of RF fields, whose maintenance dissipates power and whose injection is technologically more complex.  

However, the magnetostatic end-plug does have the drawback that the fields may not penetrate far into the rotating plasma, limiting the reflection to peripheral particles. The lowest azimuthal wavenumbers will penetrate furthest, but the averaging procedure employed here breaks down for very low mode numbers,and very low rotation frequencies.  However, to the extent that the reflection succeeds, the magnetostatic wiggler offered here can simply be used in addition to other reflection mechanisms.



\subsection*{Acknowledgments}

The authors thank Drs. I. E. Ochs, E. J. Kolmes, and M. E. Mlodik for constructive discussions. This work was supported by ARPA-E Grant No. DE-AR001554. JMR acknowledges the support and hospitality of the Princeton University Andlinger Center for Energy and the Environment. 
\section*{Author Declarations}
\subsection*{Conflict of interest}
The authors have no conflicts to disclose.
\subsection*{Data availability}
Data sharing is not applicable to this article as no new data was created or analyzed un this study.

\section*{References}
\bibliography{biblio}

\begin{thebibliography}{63}%
\makeatletter
\providecommand \@ifxundefined [1]{%
 \@ifx{#1\undefined}
}%
\providecommand \@ifnum [1]{%
 \ifnum #1\expandafter \@firstoftwo
 \else \expandafter \@secondoftwo
 \fi
}%
\providecommand \@ifx [1]{%
 \ifx #1\expandafter \@firstoftwo
 \else \expandafter \@secondoftwo
 \fi
}%
\providecommand \natexlab [1]{#1}%
\providecommand \enquote  [1]{``#1''}%
\providecommand \bibnamefont  [1]{#1}%
\providecommand \bibfnamefont [1]{#1}%
\providecommand \citenamefont [1]{#1}%
\providecommand \href@noop [0]{\@secondoftwo}%
\providecommand \href [0]{\begingroup \@sanitize@url \@href}%
\providecommand \@href[1]{\@@startlink{#1}\@@href}%
\providecommand \@@href[1]{\endgroup#1\@@endlink}%
\providecommand \@sanitize@url [0]{\catcode `\\12\catcode `\$12\catcode
  `\&12\catcode `\#12\catcode `\^12\catcode `\_12\catcode `\%12\relax}%
\providecommand \@@startlink[1]{}%
\providecommand \@@endlink[0]{}%
\providecommand \url  [0]{\begingroup\@sanitize@url \@url }%
\providecommand \@url [1]{\endgroup\@href {#1}{\urlprefix }}%
\providecommand \urlprefix  [0]{URL }%
\providecommand \Eprint [0]{\href }%
\providecommand \doibase [0]{https://doi.org/}%
\providecommand \selectlanguage [0]{\@gobble}%
\providecommand \bibinfo  [0]{\@secondoftwo}%
\providecommand \bibfield  [0]{\@secondoftwo}%
\providecommand \translation [1]{[#1]}%
\providecommand \BibitemOpen [0]{}%
\providecommand \bibitemStop [0]{}%
\providecommand \bibitemNoStop [0]{.\EOS\space}%
\providecommand \EOS [0]{\spacefactor3000\relax}%
\providecommand \BibitemShut  [1]{\csname bibitem#1\endcsname}%
\let\auto@bib@innerbib\@empty
\bibitem [{\citenamefont {Post}(1987)}]{postMagneticMirrorApproach1987}%
  \BibitemOpen
  \bibfield  {author} {\bibinfo {author} {\bibfnamefont {R.~F.}\ \bibnamefont
  {Post}},\ }\href {https://doi.org/10.1088/0029-5515/27/10/001} {\bibfield
  {journal} {\bibinfo  {journal} {Nucl. Fusion}\ }\textbf {\bibinfo {volume}
  {27}},\ \bibinfo {pages} {1579} (\bibinfo {year} {1987})}\BibitemShut
  {NoStop}%
\bibitem [{\citenamefont {Ryutov}(1988)}]{ryutovOpenendedTraps1988}%
  \BibitemOpen
  \bibfield  {author} {\bibinfo {author} {\bibfnamefont {D.~D.}\ \bibnamefont
  {Ryutov}},\ }\href {https://doi.org/10.1070/PU1988v031n04ABEH005747}
  {\bibfield  {journal} {\bibinfo  {journal} {Soviet Physics Uspekhi}\ }\textbf
  {\bibinfo {volume} {31}},\ \bibinfo {pages} {300} (\bibinfo {year}
  {1988})}\BibitemShut {NoStop}%
\bibitem [{\citenamefont {Spitzer}(1958)}]{spitzerStellaratorConcept1958}%
  \BibitemOpen
  \bibfield  {author} {\bibinfo {author} {\bibfnamefont {L.}~\bibnamefont
  {Spitzer}},\ }\href {https://doi.org/10.1063/1.1705883} {\bibfield  {journal}
  {\bibinfo  {journal} {Phys. Fluids}\ }\textbf {\bibinfo {volume} {1}},\
  \bibinfo {pages} {253} (\bibinfo {year} {1958})}\BibitemShut {NoStop}%
\bibitem [{\citenamefont {Budker}(1959)}]{budkerPlasmaPhysicsProblems1959}%
  \BibitemOpen
  \bibfield  {author} {\bibinfo {author} {\bibfnamefont {G.~I.}\ \bibnamefont
  {Budker}},\ }\href@noop {} {\emph {\bibinfo {title} {Plasma {{Physics}} and
  the {{Problems}} of {{Controlled Thermonuclear Reactions}} ({{Edited}} by
  {{MA Leontovich}})}}}\ (\bibinfo  {publisher} {{Pergamon Press, Oxford
  (IY5Y)}},\ \bibinfo {year} {1959})\BibitemShut {NoStop}%
\bibitem [{\citenamefont
  {Gormezano}(1979)}]{gormezanoReductionLossesOpenended1979}%
  \BibitemOpen
  \bibfield  {author} {\bibinfo {author} {\bibfnamefont {C.}~\bibnamefont
  {Gormezano}},\ }\href {https://doi.org/10.1088/0029-5515/19/8/008} {\bibfield
   {journal} {\bibinfo  {journal} {Nuclear Fusion}\ }\textbf {\bibinfo {volume}
  {19}},\ \bibinfo {pages} {1085} (\bibinfo {year} {1979})}\BibitemShut
  {NoStop}%
\bibitem [{\citenamefont {Baldwin}(1977)}]{baldwinEndlossProcessesMirror1977}%
  \BibitemOpen
  \bibfield  {author} {\bibinfo {author} {\bibfnamefont {D.~E.}\ \bibnamefont
  {Baldwin}},\ }\href {https://doi.org/10.1103/RevModPhys.49.317} {\bibfield
  {journal} {\bibinfo  {journal} {Reviews of Modern Physics}\ }\textbf
  {\bibinfo {volume} {49}},\ \bibinfo {pages} {317} (\bibinfo {year}
  {1977})}\BibitemShut {NoStop}%
\bibitem [{\citenamefont {Lehnert}(1971)}]{lehnertRotatingPlasmas1971}%
  \BibitemOpen
  \bibfield  {author} {\bibinfo {author} {\bibfnamefont {B.}~\bibnamefont
  {Lehnert}},\ }\href {https://doi.org/10.1088/0029-5515/11/5/010} {\bibfield
  {journal} {\bibinfo  {journal} {Nucl. Fusion}\ }\textbf {\bibinfo {volume}
  {11}},\ \bibinfo {pages} {485} (\bibinfo {year} {1971})}\BibitemShut
  {NoStop}%
\bibitem [{\citenamefont {Bekhtenev}\ \emph {et~al.}(1980)\citenamefont
  {Bekhtenev}, \citenamefont {Volosov}, \citenamefont {Pal'chikov},
  \citenamefont {Pekker},\ and\ \citenamefont
  {Yudin}}]{bekhtenevProblemsThermonuclearReactor1980}%
  \BibitemOpen
  \bibfield  {author} {\bibinfo {author} {\bibfnamefont {A.~A.}\ \bibnamefont
  {Bekhtenev}}, \bibinfo {author} {\bibfnamefont {V.~I.}\ \bibnamefont
  {Volosov}}, \bibinfo {author} {\bibfnamefont {V.~E.}\ \bibnamefont
  {Pal'chikov}}, \bibinfo {author} {\bibfnamefont {M.~S.}\ \bibnamefont
  {Pekker}},\ and\ \bibinfo {author} {\bibfnamefont {Y.~N.}\ \bibnamefont
  {Yudin}},\ }\href {https://doi.org/10.1088/0029-5515/20/5/007} {\bibfield
  {journal} {\bibinfo  {journal} {Nuclear Fusion}\ }\textbf {\bibinfo {volume}
  {20}},\ \bibinfo {pages} {579} (\bibinfo {year} {1980})}\BibitemShut
  {NoStop}%
\bibitem [{\citenamefont {Volosov}\ and\ \citenamefont
  {Pekker}(1981)}]{volosovLongitudinalPlasmaConfinement1981}%
  \BibitemOpen
  \bibfield  {author} {\bibinfo {author} {\bibfnamefont {V.}~\bibnamefont
  {Volosov}}\ and\ \bibinfo {author} {\bibfnamefont {M.}~\bibnamefont
  {Pekker}},\ }\href {https://doi.org/10.1088/0029-5515/21/10/006} {\bibfield
  {journal} {\bibinfo  {journal} {Nuclear Fusion}\ }\textbf {\bibinfo {volume}
  {21}},\ \bibinfo {pages} {1275} (\bibinfo {year} {1981})}\BibitemShut
  {NoStop}%
\bibitem [{\citenamefont {Hassam}(1997)}]{hassamSteadyStateCentrifugally1997}%
  \BibitemOpen
  \bibfield  {author} {\bibinfo {author} {\bibfnamefont {A.~B.}\ \bibnamefont
  {Hassam}},\ }\href@noop {} {\bibfield  {journal} {\bibinfo  {journal}
  {Comments on Plasma Physics and Controlled Fusion}\ }\textbf {\bibinfo
  {volume} {18}},\ \bibinfo {pages} {263} (\bibinfo {year} {1997})}\BibitemShut
  {NoStop}%
\bibitem [{\citenamefont {Teodorescu}\ \emph {et~al.}(2010)\citenamefont
  {Teodorescu}, \citenamefont {Young}, \citenamefont {Swan}, \citenamefont
  {Ellis}, \citenamefont {Hassam},\ and\ \citenamefont
  {{Romero-Talamas}}}]{teodorescuConfinementPlasmaShaped2010}%
  \BibitemOpen
  \bibfield  {author} {\bibinfo {author} {\bibfnamefont {C.}~\bibnamefont
  {Teodorescu}}, \bibinfo {author} {\bibfnamefont {W.~C.}\ \bibnamefont
  {Young}}, \bibinfo {author} {\bibfnamefont {G.~W.~S.}\ \bibnamefont {Swan}},
  \bibinfo {author} {\bibfnamefont {R.~F.}\ \bibnamefont {Ellis}}, \bibinfo
  {author} {\bibfnamefont {A.~B.}\ \bibnamefont {Hassam}},\ and\ \bibinfo
  {author} {\bibfnamefont {C.~A.}\ \bibnamefont {{Romero-Talamas}}},\ }\href
  {https://doi.org/10.1103/PhysRevLett.105.085003} {\bibfield  {journal}
  {\bibinfo  {journal} {Phys. Rev. Lett.}\ }\textbf {\bibinfo {volume} {105}},\
  \bibinfo {pages} {085003} (\bibinfo {year} {2010})}\BibitemShut {NoStop}%
\bibitem [{\citenamefont {Fetterman}\ and\ \citenamefont
  {Fisch}(2010{\natexlab{a}})}]{fettermanAlphaChannelingRotating2010}%
  \BibitemOpen
  \bibfield  {author} {\bibinfo {author} {\bibfnamefont {A.~J.}\ \bibnamefont
  {Fetterman}}\ and\ \bibinfo {author} {\bibfnamefont {N.~J.}\ \bibnamefont
  {Fisch}},\ }\href {https://doi.org/10.1063/1.3389308} {\bibfield  {journal}
  {\bibinfo  {journal} {Phys. Plasmas}\ }\textbf {\bibinfo {volume} {17}},\
  \bibinfo {pages} {042112} (\bibinfo {year} {2010}{\natexlab{a}})}\BibitemShut
  {NoStop}%
\bibitem [{\citenamefont {Fetterman}\ and\ \citenamefont
  {Fisch}(2010{\natexlab{b}})}]{fettermanWavedrivenRotationSupersonically2010}%
  \BibitemOpen
  \bibfield  {author} {\bibinfo {author} {\bibfnamefont {A.~J.}\ \bibnamefont
  {Fetterman}}\ and\ \bibinfo {author} {\bibfnamefont {N.~J.}\ \bibnamefont
  {Fisch}},\ }\href {https://doi.org/10.13182/FST10-A9496} {\bibfield
  {journal} {\bibinfo  {journal} {Fusion Sci. Technol.}\ }\textbf {\bibinfo
  {volume} {57}},\ \bibinfo {pages} {343} (\bibinfo {year}
  {2010}{\natexlab{b}})}\BibitemShut {NoStop}%
\bibitem [{\citenamefont {Fetterman}\ and\ \citenamefont
  {Fisch}(2008)}]{fettermanChannelingRotatingPlasma2008}%
  \BibitemOpen
  \bibfield  {author} {\bibinfo {author} {\bibfnamefont {A.~J.}\ \bibnamefont
  {Fetterman}}\ and\ \bibinfo {author} {\bibfnamefont {N.~J.}\ \bibnamefont
  {Fisch}},\ }\href {https://doi.org/10.1103/PhysRevLett.101.205003} {\bibfield
   {journal} {\bibinfo  {journal} {Phys. Rev. Lett.}\ }\textbf {\bibinfo
  {volume} {101}},\ \bibinfo {pages} {205003} (\bibinfo {year}
  {2008})}\BibitemShut {NoStop}%
\bibitem [{\citenamefont {Fowler}\ \emph {et~al.}(2017)\citenamefont {Fowler},
  \citenamefont {Moir},\ and\ \citenamefont
  {Simonen}}]{fowlerNewSimplerWay2017}%
  \BibitemOpen
  \bibfield  {author} {\bibinfo {author} {\bibfnamefont {T.}~\bibnamefont
  {Fowler}}, \bibinfo {author} {\bibfnamefont {R.}~\bibnamefont {Moir}},\ and\
  \bibinfo {author} {\bibfnamefont {T.}~\bibnamefont {Simonen}},\ }\href
  {https://doi.org/10.1088/1741-4326/aa5e54} {\bibfield  {journal} {\bibinfo
  {journal} {Nuclear Fusion}\ }\textbf {\bibinfo {volume} {57}},\ \bibinfo
  {pages} {056014} (\bibinfo {year} {2017})}\BibitemShut {NoStop}%
\bibitem [{\citenamefont {White}\ \emph {et~al.}(2018)\citenamefont {White},
  \citenamefont {Hassam},\ and\ \citenamefont
  {Brizard}}]{whiteCentrifugalParticleConfinement2018}%
  \BibitemOpen
  \bibfield  {author} {\bibinfo {author} {\bibfnamefont {R.}~\bibnamefont
  {White}}, \bibinfo {author} {\bibfnamefont {A.}~\bibnamefont {Hassam}},\ and\
  \bibinfo {author} {\bibfnamefont {A.}~\bibnamefont {Brizard}},\ }\href
  {https://doi.org/10.1063/1.5003359} {\bibfield  {journal} {\bibinfo
  {journal} {Physics of Plasmas}\ }\textbf {\bibinfo {volume} {25}},\ \bibinfo
  {pages} {012514} (\bibinfo {year} {2018})}\BibitemShut {NoStop}%
\bibitem [{\citenamefont
  {Beklemishev}(2013)}]{beklemishevHelicoidalSystemAxial2013}%
  \BibitemOpen
  \bibfield  {author} {\bibinfo {author} {\bibfnamefont {A.~D.}\ \bibnamefont
  {Beklemishev}},\ }\href {https://doi.org/10.13182/FST13-A16953} {\bibfield
  {journal} {\bibinfo  {journal} {Fusion Science and Technology}\ }\textbf
  {\bibinfo {volume} {63}},\ \bibinfo {pages} {355} (\bibinfo {year}
  {2013})}\BibitemShut {NoStop}%
\bibitem [{\citenamefont
  {Beklemishev}(2015)}]{beklemishevHelicalPlasmaThruster2015}%
  \BibitemOpen
  \bibfield  {author} {\bibinfo {author} {\bibfnamefont {A.~D.}\ \bibnamefont
  {Beklemishev}},\ }\href {https://doi.org/10.1063/1.4932075} {\bibfield
  {journal} {\bibinfo  {journal} {Physics of Plasmas}\ }\textbf {\bibinfo
  {volume} {22}},\ \bibinfo {pages} {103506} (\bibinfo {year}
  {2015})}\BibitemShut {NoStop}%
\bibitem [{\citenamefont {Postupaev}\ \emph {et~al.}(2016)\citenamefont
  {Postupaev}, \citenamefont {Sudnikov}, \citenamefont {Beklemishev},\ and\
  \citenamefont {Ivanov}}]{postupaevHelicalMirrorsActive2016}%
  \BibitemOpen
  \bibfield  {author} {\bibinfo {author} {\bibfnamefont {V.}~\bibnamefont
  {Postupaev}}, \bibinfo {author} {\bibfnamefont {A.}~\bibnamefont {Sudnikov}},
  \bibinfo {author} {\bibfnamefont {A.}~\bibnamefont {Beklemishev}},\ and\
  \bibinfo {author} {\bibfnamefont {I.}~\bibnamefont {Ivanov}},\ }\href
  {https://doi.org/10.1016/j.fusengdes.2016.03.029} {\bibfield  {journal}
  {\bibinfo  {journal} {Fusion Engineering and Design}\ }\textbf {\bibinfo
  {volume} {106}},\ \bibinfo {pages} {29} (\bibinfo {year} {2016})}\BibitemShut
  {NoStop}%
\bibitem [{\citenamefont {Anderegg}\ \emph {et~al.}(1995)\citenamefont
  {Anderegg}, \citenamefont {Huang}, \citenamefont {Driscoll}, \citenamefont
  {Severn},\ and\ \citenamefont {Sarid}}]{andereggLongIonPlasma1995}%
  \BibitemOpen
  \bibfield  {author} {\bibinfo {author} {\bibfnamefont {F.}~\bibnamefont
  {Anderegg}}, \bibinfo {author} {\bibfnamefont {X.-P.}\ \bibnamefont {Huang}},
  \bibinfo {author} {\bibfnamefont {C.~F.}\ \bibnamefont {Driscoll}}, \bibinfo
  {author} {\bibfnamefont {G.~D.}\ \bibnamefont {Severn}},\ and\ \bibinfo
  {author} {\bibfnamefont {E.}~\bibnamefont {Sarid}},\ }in\ \href
  {https://doi.org/10.1063/1.47893} {\emph {\bibinfo {booktitle} {{{AIP
  Conference Proceedings}}}}},\ Vol.\ \bibinfo {volume} {331}\ (\bibinfo
  {publisher} {{AIP}},\ \bibinfo {address} {{Berkeley, California (USA) plasmas
  in traps}},\ \bibinfo {year} {1995})\ pp.\ \bibinfo {pages}
  {1--6}\BibitemShut {NoStop}%
\bibitem [{\citenamefont {Gueroult}\ \emph {et~al.}(2015)\citenamefont
  {Gueroult}, \citenamefont {Hobbs},\ and\ \citenamefont
  {Fisch}}]{gueroultPlasmaFilteringTechniques2015}%
  \BibitemOpen
  \bibfield  {author} {\bibinfo {author} {\bibfnamefont {R.}~\bibnamefont
  {Gueroult}}, \bibinfo {author} {\bibfnamefont {D.~T.}\ \bibnamefont
  {Hobbs}},\ and\ \bibinfo {author} {\bibfnamefont {N.~J.}\ \bibnamefont
  {Fisch}},\ }\href {https://doi.org/10.1016/j.jhazmat.2015.04.058} {\bibfield
  {journal} {\bibinfo  {journal} {J. Hazard. Mater.}\ }\textbf {\bibinfo
  {volume} {297}},\ \bibinfo {pages} {153} (\bibinfo {year}
  {2015})}\BibitemShut {NoStop}%
\bibitem [{\citenamefont {Dolgolenko}\ and\ \citenamefont {{Yu. A.
  Muromkin}}(2017)}]{dolgolenkoSeparationMixturesChemical2017}%
  \BibitemOpen
  \bibfield  {author} {\bibinfo {author} {\bibfnamefont {D.~A.}\ \bibnamefont
  {Dolgolenko}}\ and\ \bibinfo {author} {\bibnamefont {{Yu. A. Muromkin}}},\
  }\href {https://doi.org/10.3367/UFNe.2016.12.038016} {\bibfield  {journal}
  {\bibinfo  {journal} {Phys.-Usp.}\ }\textbf {\bibinfo {volume} {60}},\
  \bibinfo {pages} {994} (\bibinfo {year} {2017})}\BibitemShut {NoStop}%
\bibitem [{\citenamefont
  {Timofeev}(2014)}]{timofeevTheoryPlasmaProcessing2014}%
  \BibitemOpen
  \bibfield  {author} {\bibinfo {author} {\bibfnamefont {A.~V.}\ \bibnamefont
  {Timofeev}},\ }\href {https://doi.org/10.3367/UFNe.0184.201410g.1101}
  {\bibfield  {journal} {\bibinfo  {journal} {Physics-Uspekhi}\ }\textbf
  {\bibinfo {volume} {57}},\ \bibinfo {pages} {990} (\bibinfo {year}
  {2014})}\BibitemShut {NoStop}%
\bibitem [{\citenamefont {Gueroult}\ and\ \citenamefont
  {Fisch}(2014)}]{gueroultPlasmaMassFiltering2014}%
  \BibitemOpen
  \bibfield  {author} {\bibinfo {author} {\bibfnamefont {R.}~\bibnamefont
  {Gueroult}}\ and\ \bibinfo {author} {\bibfnamefont {N.~J.}\ \bibnamefont
  {Fisch}},\ }\href {https://doi.org/10.1088/0963-0252/23/3/035002} {\bibfield
  {journal} {\bibinfo  {journal} {Plasma Sources Sci. Technol.}\ }\textbf
  {\bibinfo {volume} {23}},\ \bibinfo {pages} {035002} (\bibinfo {year}
  {2014})}\BibitemShut {NoStop}%
\bibitem [{\citenamefont {Vorona}\ \emph {et~al.}(2015)\citenamefont {Vorona},
  \citenamefont {Gavrikov}, \citenamefont {Samokhin}, \citenamefont {Smirnov},\
  and\ \citenamefont {Khomyakov}}]{voronaPossibilityReprocessingSpent2015}%
  \BibitemOpen
  \bibfield  {author} {\bibinfo {author} {\bibfnamefont {N.~A.}\ \bibnamefont
  {Vorona}}, \bibinfo {author} {\bibfnamefont {A.~V.}\ \bibnamefont
  {Gavrikov}}, \bibinfo {author} {\bibfnamefont {A.~A.}\ \bibnamefont
  {Samokhin}}, \bibinfo {author} {\bibfnamefont {V.~P.}\ \bibnamefont
  {Smirnov}},\ and\ \bibinfo {author} {\bibfnamefont {Y.~S.}\ \bibnamefont
  {Khomyakov}},\ }\href {https://doi.org/10.1134/S1063778815140148} {\bibfield
  {journal} {\bibinfo  {journal} {Physics of Atomic Nuclei}\ }\textbf {\bibinfo
  {volume} {78}},\ \bibinfo {pages} {1624} (\bibinfo {year}
  {2015})}\BibitemShut {NoStop}%
\bibitem [{\citenamefont {Litvak}\ \emph {et~al.}(2003)\citenamefont {Litvak},
  \citenamefont {Agnew}, \citenamefont {Anderegg}, \citenamefont {Cluggish},
  \citenamefont {Freeman}, \citenamefont {Gilleland}, \citenamefont {Isler},
  \citenamefont {Lee}, \citenamefont {Miller},\ and\ \citenamefont
  {Ohkawa}}]{litvakArchimedesPlasmaMass2003}%
  \BibitemOpen
  \bibfield  {author} {\bibinfo {author} {\bibfnamefont {A.}~\bibnamefont
  {Litvak}}, \bibinfo {author} {\bibfnamefont {S.}~\bibnamefont {Agnew}},
  \bibinfo {author} {\bibfnamefont {F.}~\bibnamefont {Anderegg}}, \bibinfo
  {author} {\bibfnamefont {B.}~\bibnamefont {Cluggish}}, \bibinfo {author}
  {\bibfnamefont {R.}~\bibnamefont {Freeman}}, \bibinfo {author} {\bibfnamefont
  {J.}~\bibnamefont {Gilleland}}, \bibinfo {author} {\bibfnamefont
  {R.}~\bibnamefont {Isler}}, \bibinfo {author} {\bibfnamefont
  {W.}~\bibnamefont {Lee}}, \bibinfo {author} {\bibfnamefont {R.}~\bibnamefont
  {Miller}},\ and\ \bibinfo {author} {\bibfnamefont {T.}~\bibnamefont
  {Ohkawa}},\ }in\ \href@noop {} {\emph {\bibinfo {booktitle} {30th {{EPS
  Conference}} on {{Contr}}. {{Fusion}} and {{Plasma Phys}}}}},\ Vol.~\bibinfo
  {volume} {27}\ (\bibinfo {year} {2003})\BibitemShut {NoStop}%
\bibitem [{\citenamefont
  {Janes}(1965)}]{janesExperimentsMagneticallyProduced1965}%
  \BibitemOpen
  \bibfield  {author} {\bibinfo {author} {\bibfnamefont {G.~S.}\ \bibnamefont
  {Janes}},\ }\href {https://doi.org/10.1103/PhysRevLett.15.135} {\bibfield
  {journal} {\bibinfo  {journal} {Phys. Rev. Lett.}\ }\textbf {\bibinfo
  {volume} {15}},\ \bibinfo {pages} {135} (\bibinfo {year} {1965})}\BibitemShut
  {NoStop}%
\bibitem [{\citenamefont {Janes}\ \emph {et~al.}(1966)\citenamefont {Janes},
  \citenamefont {Levy}, \citenamefont {Bethe},\ and\ \citenamefont
  {Field}}]{janesNewTypeAccelerator1966}%
  \BibitemOpen
  \bibfield  {author} {\bibinfo {author} {\bibfnamefont {G.~S.}\ \bibnamefont
  {Janes}}, \bibinfo {author} {\bibfnamefont {R.~H.}\ \bibnamefont {Levy}},
  \bibinfo {author} {\bibfnamefont {H.~A.}\ \bibnamefont {Bethe}},\ and\
  \bibinfo {author} {\bibfnamefont {B.~T.}\ \bibnamefont {Field}},\ }\href
  {https://doi.org/10.1103/PhysRev.145.925} {\bibfield  {journal} {\bibinfo
  {journal} {Phys. Rev.}\ }\textbf {\bibinfo {volume} {145}},\ \bibinfo {pages}
  {925} (\bibinfo {year} {1966})}\BibitemShut {NoStop}%
\bibitem [{\citenamefont {Janes}\ \emph {et~al.}(1965)\citenamefont {Janes},
  \citenamefont {Levy},\ and\ \citenamefont
  {Petschek}}]{janesProductionBeVPotential1965}%
  \BibitemOpen
  \bibfield  {author} {\bibinfo {author} {\bibfnamefont {G.~S.}\ \bibnamefont
  {Janes}}, \bibinfo {author} {\bibfnamefont {R.~H.}\ \bibnamefont {Levy}},\
  and\ \bibinfo {author} {\bibfnamefont {H.~E.}\ \bibnamefont {Petschek}},\
  }\href {https://doi.org/10.1103/PhysRevLett.15.138} {\bibfield  {journal}
  {\bibinfo  {journal} {Phys. Rev. Lett.}\ }\textbf {\bibinfo {volume} {15}},\
  \bibinfo {pages} {138} (\bibinfo {year} {1965})}\BibitemShut {NoStop}%
\bibitem [{\citenamefont {Rax}\ \emph {et~al.}(2017)\citenamefont {Rax},
  \citenamefont {Gueroult},\ and\ \citenamefont
  {Fisch}}]{raxEfficiencyWavedrivenRigid2017}%
  \BibitemOpen
  \bibfield  {author} {\bibinfo {author} {\bibfnamefont {J.-M.}\ \bibnamefont
  {Rax}}, \bibinfo {author} {\bibfnamefont {R.}~\bibnamefont {Gueroult}},\ and\
  \bibinfo {author} {\bibfnamefont {N.~J.}\ \bibnamefont {Fisch}},\ }\href
  {https://doi.org/10.1063/1.4977919} {\bibfield  {journal} {\bibinfo
  {journal} {Phys. Plasmas}\ }\textbf {\bibinfo {volume} {24}},\ \bibinfo
  {pages} {032504} (\bibinfo {year} {2017})}\BibitemShut {NoStop}%
\bibitem [{\citenamefont {Ochs}\ and\ \citenamefont
  {Fisch}(2017)}]{ochsParticleOrbitsForcebalanced2017}%
  \BibitemOpen
  \bibfield  {author} {\bibinfo {author} {\bibfnamefont {I.~E.}\ \bibnamefont
  {Ochs}}\ and\ \bibinfo {author} {\bibfnamefont {N.~J.}\ \bibnamefont
  {Fisch}},\ }\href {https://doi.org/10.1063/1.4991510} {\bibfield  {journal}
  {\bibinfo  {journal} {Phys. Plasmas}\ }\textbf {\bibinfo {volume} {24}},\
  \bibinfo {pages} {092513} (\bibinfo {year} {2017})}\BibitemShut {NoStop}%
\bibitem [{\citenamefont {Motz}\ and\ \citenamefont
  {Watson}(1967)}]{motzRadioFrequencyConfinementAcceleration1967}%
  \BibitemOpen
  \bibfield  {author} {\bibinfo {author} {\bibfnamefont {H.}~\bibnamefont
  {Motz}}\ and\ \bibinfo {author} {\bibfnamefont {C.}~\bibnamefont {Watson}},\
  }in\ \href {https://doi.org/10.1016/S0065-2539(08)60061-X} {\emph {\bibinfo
  {booktitle} {Advances in {{Electronics}} and {{Electron Physics}}}}},\
  Vol.~\bibinfo {volume} {23}\ (\bibinfo  {publisher} {{Elsevier}},\ \bibinfo
  {year} {1967})\ pp.\ \bibinfo {pages} {153--302}\BibitemShut {NoStop}%
\bibitem [{\citenamefont {Gaponov}\ and\ \citenamefont
  {Miller}(1958)}]{gaponov1958potential}%
  \BibitemOpen
  \bibfield  {author} {\bibinfo {author} {\bibfnamefont {A.}~\bibnamefont
  {Gaponov}}\ and\ \bibinfo {author} {\bibfnamefont {M.}~\bibnamefont
  {Miller}},\ }\href@noop {} {\bibfield  {journal} {\bibinfo  {journal}
  {Journal of Experimental and Theoretical Physics}\ }\textbf {\bibinfo
  {volume} {34}},\ \bibinfo {pages} {242} (\bibinfo {year} {1958})}\BibitemShut
  {NoStop}%
\bibitem [{\citenamefont {Moir}\ and\ \citenamefont
  {Barr}(1973)}]{moirVenetianblindDirectEnergy1973}%
  \BibitemOpen
  \bibfield  {author} {\bibinfo {author} {\bibfnamefont {R.}~\bibnamefont
  {Moir}}\ and\ \bibinfo {author} {\bibfnamefont {W.}~\bibnamefont {Barr}},\
  }\href {https://doi.org/10.1088/0029-5515/13/1/005} {\bibfield  {journal}
  {\bibinfo  {journal} {Nuclear Fusion}\ }\textbf {\bibinfo {volume} {13}},\
  \bibinfo {pages} {35} (\bibinfo {year} {1973})}\BibitemShut {NoStop}%
\bibitem [{\citenamefont {Miley}(1976)}]{mileyFusionEnergyConversion1976}%
  \BibitemOpen
  \bibfield  {author} {\bibinfo {author} {\bibfnamefont {G.~H.}\ \bibnamefont
  {Miley}},\ }\href@noop {} {\bibfield  {journal} {\bibinfo  {journal} {Amer
  Nuclear Society}\ } (\bibinfo {year} {1976})}\BibitemShut {NoStop}%
\bibitem [{\citenamefont {Taniguchi}\ \emph {et~al.}(2010)\citenamefont
  {Taniguchi}, \citenamefont {Sotani}, \citenamefont {Yasaka},\ and\
  \citenamefont {Takeno}}]{taniguchiStudiesChargeSeparation2010}%
  \BibitemOpen
  \bibfield  {author} {\bibinfo {author} {\bibfnamefont {A.}~\bibnamefont
  {Taniguchi}}, \bibinfo {author} {\bibfnamefont {N.}~\bibnamefont {Sotani}},
  \bibinfo {author} {\bibfnamefont {Y.}~\bibnamefont {Yasaka}},\ and\ \bibinfo
  {author} {\bibfnamefont {H.}~\bibnamefont {Takeno}},\ }\href@noop {}
  {\bibfield  {journal} {\bibinfo  {journal} {J. Plasma and Fusion Res.
  Series}\ } (\bibinfo {year} {2010})}\BibitemShut {NoStop}%
\bibitem [{\citenamefont {Takeno}\ \emph {et~al.}(2019)\citenamefont {Takeno},
  \citenamefont {Ichimura}, \citenamefont {Nakamoto}, \citenamefont
  {Nakashima}, \citenamefont {Matsuura}, \citenamefont {Miyazawa},
  \citenamefont {Goto}, \citenamefont {Furuyama},\ and\ \citenamefont
  {Taniike}}]{takenoRecentAdvancementResearch2019}%
  \BibitemOpen
  \bibfield  {author} {\bibinfo {author} {\bibfnamefont {H.}~\bibnamefont
  {Takeno}}, \bibinfo {author} {\bibfnamefont {K.}~\bibnamefont {Ichimura}},
  \bibinfo {author} {\bibfnamefont {S.}~\bibnamefont {Nakamoto}}, \bibinfo
  {author} {\bibfnamefont {Y.}~\bibnamefont {Nakashima}}, \bibinfo {author}
  {\bibfnamefont {H.}~\bibnamefont {Matsuura}}, \bibinfo {author}
  {\bibfnamefont {J.}~\bibnamefont {Miyazawa}}, \bibinfo {author}
  {\bibfnamefont {T.}~\bibnamefont {Goto}}, \bibinfo {author} {\bibfnamefont
  {Y.}~\bibnamefont {Furuyama}},\ and\ \bibinfo {author} {\bibfnamefont
  {A.}~\bibnamefont {Taniike}},\ }\href
  {https://doi.org/10.1585/pfr.14.2405013} {\bibfield  {journal} {\bibinfo
  {journal} {Plasma and Fusion Research}\ }\textbf {\bibinfo {volume} {14}},\
  \bibinfo {pages} {2405013} (\bibinfo {year} {2019})}\BibitemShut {NoStop}%
\bibitem [{\citenamefont
  {Volosov}(2005)}]{volosovRecuperationChargedParticle2005}%
  \BibitemOpen
  \bibfield  {author} {\bibinfo {author} {\bibfnamefont {V.~I.}\ \bibnamefont
  {Volosov}},\ }\href {https://doi.org/10.13182/FST05-A687} {\bibfield
  {journal} {\bibinfo  {journal} {Fusion Science and Technology}\ }\textbf
  {\bibinfo {volume} {47}},\ \bibinfo {pages} {351} (\bibinfo {year}
  {2005})}\BibitemShut {NoStop}%
\bibitem [{\citenamefont {Cary}\ and\ \citenamefont
  {Kaufman}(1977)}]{caryPonderomotiveForceLinear1977}%
  \BibitemOpen
  \bibfield  {author} {\bibinfo {author} {\bibfnamefont {J.~R.}\ \bibnamefont
  {Cary}}\ and\ \bibinfo {author} {\bibfnamefont {A.~N.}\ \bibnamefont
  {Kaufman}},\ }\href {https://doi.org/10.1103/PhysRevLett.39.402} {\bibfield
  {journal} {\bibinfo  {journal} {Physical Review Letters}\ }\textbf {\bibinfo
  {volume} {39}},\ \bibinfo {pages} {402} (\bibinfo {year} {1977})}\BibitemShut
  {NoStop}%
\bibitem [{\citenamefont {Dimonte}\ \emph {et~al.}(1983)\citenamefont
  {Dimonte}, \citenamefont {Lamb},\ and\ \citenamefont
  {Morales}}]{dimonteEffectsNonadiabaticityApplications1983}%
  \BibitemOpen
  \bibfield  {author} {\bibinfo {author} {\bibfnamefont {G.}~\bibnamefont
  {Dimonte}}, \bibinfo {author} {\bibfnamefont {B.~M.}\ \bibnamefont {Lamb}},\
  and\ \bibinfo {author} {\bibfnamefont {G.~J.}\ \bibnamefont {Morales}},\
  }\href {https://doi.org/10.1088/0032-1028/25/7/002} {\bibfield  {journal}
  {\bibinfo  {journal} {Plasma Physics}\ }\textbf {\bibinfo {volume} {25}},\
  \bibinfo {pages} {713} (\bibinfo {year} {1983})}\BibitemShut {NoStop}%
\bibitem [{\citenamefont {Kono}\ and\ \citenamefont
  {Sanuki}(1987)}]{konoPonderomotiveForceCyclotron1987}%
  \BibitemOpen
  \bibfield  {author} {\bibinfo {author} {\bibfnamefont {M.}~\bibnamefont
  {Kono}}\ and\ \bibinfo {author} {\bibfnamefont {H.}~\bibnamefont {Sanuki}},\
  }\href {https://doi.org/10.1017/S0022377800012393} {\bibfield  {journal}
  {\bibinfo  {journal} {Journal of Plasma Physics}\ }\textbf {\bibinfo {volume}
  {38}},\ \bibinfo {pages} {43} (\bibinfo {year} {1987})}\BibitemShut {NoStop}%
\bibitem [{\citenamefont {Grossman}\ \emph {et~al.}(1992)\citenamefont
  {Grossman}, \citenamefont {Schmitz}, \citenamefont {Najmabadi},\ and\
  \citenamefont {Conn}}]{grossmanRFPonderomotiveForces1992}%
  \BibitemOpen
  \bibfield  {author} {\bibinfo {author} {\bibfnamefont {A.}~\bibnamefont
  {Grossman}}, \bibinfo {author} {\bibfnamefont {L.}~\bibnamefont {Schmitz}},
  \bibinfo {author} {\bibfnamefont {F.}~\bibnamefont {Najmabadi}},\ and\
  \bibinfo {author} {\bibfnamefont {R.}~\bibnamefont {Conn}},\ }\href
  {https://doi.org/10.1016/S0022-3115(06)80141-8} {\bibfield  {journal}
  {\bibinfo  {journal} {Journal of Nuclear Materials}\ }\textbf {\bibinfo
  {volume} {196--198}},\ \bibinfo {pages} {775} (\bibinfo {year}
  {1992})}\BibitemShut {NoStop}%
\bibitem [{\citenamefont {Masuzaki}\ \emph {et~al.}(1995)\citenamefont
  {Masuzaki}, \citenamefont {Ohno},\ and\ \citenamefont
  {Takamura}}]{masuzakiReductionPlasmaHeat1995}%
  \BibitemOpen
  \bibfield  {author} {\bibinfo {author} {\bibfnamefont {S.}~\bibnamefont
  {Masuzaki}}, \bibinfo {author} {\bibfnamefont {N.}~\bibnamefont {Ohno}},\
  and\ \bibinfo {author} {\bibfnamefont {S.}~\bibnamefont {Takamura}},\ }\href
  {https://doi.org/10.1016/0022-3115(94)00484-6} {\bibfield  {journal}
  {\bibinfo  {journal} {Journal of Nuclear Materials}\ }\textbf {\bibinfo
  {volume} {220--222}},\ \bibinfo {pages} {1112} (\bibinfo {year}
  {1995})}\BibitemShut {NoStop}%
\bibitem [{\citenamefont {Tokman}(1999)}]{tokman99}%
  \BibitemOpen
  \bibfield  {author} {\bibinfo {author} {\bibfnamefont {M.~D.}\ \bibnamefont
  {Tokman}},\ }\href@noop {} {\bibfield  {journal} {\bibinfo  {journal} {Fizika
  Plazmy}\ }\textbf {\bibinfo {volume} {25}},\ \bibinfo {pages} {160} (\bibinfo
  {year} {1999})}\BibitemShut {NoStop}%
\bibitem [{\citenamefont {Dodin}\ and\ \citenamefont
  {Fisch}(2005{\natexlab{a}})}]{dodinPonderomotiveRatchetUniform2005}%
  \BibitemOpen
  \bibfield  {author} {\bibinfo {author} {\bibfnamefont {I.~Y.}\ \bibnamefont
  {Dodin}}\ and\ \bibinfo {author} {\bibfnamefont {N.~J.}\ \bibnamefont
  {Fisch}},\ }\href {https://doi.org/10.1103/PhysRevE.72.046602} {\bibfield
  {journal} {\bibinfo  {journal} {Phys. Rev. E}\ }\textbf {\bibinfo {volume}
  {72}},\ \bibinfo {pages} {046602} (\bibinfo {year}
  {2005}{\natexlab{a}})}\BibitemShut {NoStop}%
\bibitem [{\citenamefont {Dodin}\ and\ \citenamefont
  {Fisch}(2005{\natexlab{b}})}]{dodinQuantumlikeDynamicsClassical2005}%
  \BibitemOpen
  \bibfield  {author} {\bibinfo {author} {\bibfnamefont {I.~Y.}\ \bibnamefont
  {Dodin}}\ and\ \bibinfo {author} {\bibfnamefont {N.~J.}\ \bibnamefont
  {Fisch}},\ }\href {https://doi.org/10.1103/PhysRevLett.95.115001} {\bibfield
  {journal} {\bibinfo  {journal} {Physical Review Letters}\ }\textbf {\bibinfo
  {volume} {95}},\ \bibinfo {pages} {115001} (\bibinfo {year}
  {2005}{\natexlab{b}})}\BibitemShut {NoStop}%
\bibitem [{\citenamefont {Dodin}\ and\ \citenamefont
  {Fisch}(2006{\natexlab{a}})}]{dodinNonadiabaticPonderomotivePotentials2006}%
  \BibitemOpen
  \bibfield  {author} {\bibinfo {author} {\bibfnamefont {I.}~\bibnamefont
  {Dodin}}\ and\ \bibinfo {author} {\bibfnamefont {N.}~\bibnamefont {Fisch}},\
  }\href {https://doi.org/10.1016/j.physleta.2005.09.049} {\bibfield  {journal}
  {\bibinfo  {journal} {Physics Letters A}\ }\textbf {\bibinfo {volume}
  {349}},\ \bibinfo {pages} {356} (\bibinfo {year}
  {2006}{\natexlab{a}})}\BibitemShut {NoStop}%
\bibitem [{\citenamefont {Dodin}\ and\ \citenamefont
  {Fisch}(2006{\natexlab{b}})}]{dodinNonadiabaticTunnelingPonderomotive2006}%
  \BibitemOpen
  \bibfield  {author} {\bibinfo {author} {\bibfnamefont {I.~Y.}\ \bibnamefont
  {Dodin}}\ and\ \bibinfo {author} {\bibfnamefont {N.~J.}\ \bibnamefont
  {Fisch}},\ }\href {https://doi.org/10.1103/PhysRevE.74.056404} {\bibfield
  {journal} {\bibinfo  {journal} {Physical Review E}\ }\textbf {\bibinfo
  {volume} {74}},\ \bibinfo {pages} {056404} (\bibinfo {year}
  {2006}{\natexlab{b}})}\BibitemShut {NoStop}%
\bibitem [{\citenamefont {Suvorov}\ and\ \citenamefont
  {Tokman}(1988)}]{suvorov88}%
  \BibitemOpen
  \bibfield  {author} {\bibinfo {author} {\bibfnamefont {E.~V.}\ \bibnamefont
  {Suvorov}}\ and\ \bibinfo {author} {\bibfnamefont {M.~D.}\ \bibnamefont
  {Tokman}},\ }\href@noop {} {\bibfield  {journal} {\bibinfo  {journal} {Fizika
  Plazmy}\ }\textbf {\bibinfo {volume} {14}},\ \bibinfo {pages} {950} (\bibinfo
  {year} {1988})}\BibitemShut {NoStop}%
\bibitem [{\citenamefont {Litvak}\ \emph {et~al.}(1993)\citenamefont {Litvak},
  \citenamefont {Sergeev}, \citenamefont {Suvorov}, \citenamefont {Tokman},\
  and\ \citenamefont {Khazanov}}]{litvakNonlinearEffectsElectron1993}%
  \BibitemOpen
  \bibfield  {author} {\bibinfo {author} {\bibfnamefont {A.~G.}\ \bibnamefont
  {Litvak}}, \bibinfo {author} {\bibfnamefont {A.~M.}\ \bibnamefont {Sergeev}},
  \bibinfo {author} {\bibfnamefont {E.~V.}\ \bibnamefont {Suvorov}}, \bibinfo
  {author} {\bibfnamefont {M.~D.}\ \bibnamefont {Tokman}},\ and\ \bibinfo
  {author} {\bibfnamefont {I.~V.}\ \bibnamefont {Khazanov}},\ }\href
  {https://doi.org/10.1063/1.860552} {\bibfield  {journal} {\bibinfo  {journal}
  {Physics of Fluids B: Plasma Physics}\ }\textbf {\bibinfo {volume} {5}},\
  \bibinfo {pages} {4347} (\bibinfo {year} {1993})}\BibitemShut {NoStop}%
\bibitem [{\citenamefont {Fisch}\ \emph {et~al.}(2003)\citenamefont {Fisch},
  \citenamefont {Rax},\ and\ \citenamefont
  {Dodin}}]{fischCurrentDrivePonderomotive2003}%
  \BibitemOpen
  \bibfield  {author} {\bibinfo {author} {\bibfnamefont {N.~J.}\ \bibnamefont
  {Fisch}}, \bibinfo {author} {\bibfnamefont {J.~M.}\ \bibnamefont {Rax}},\
  and\ \bibinfo {author} {\bibfnamefont {I.~Y.}\ \bibnamefont {Dodin}},\ }\href
  {https://doi.org/10.1103/PhysRevLett.91.205004} {\bibfield  {journal}
  {\bibinfo  {journal} {Physical Review Letters}\ }\textbf {\bibinfo {volume}
  {91}},\ \bibinfo {pages} {205004} (\bibinfo {year} {2003})},\ \bibinfo {note}
  {\textit{ibid} {\bf 93}, 059902(E) (2004).}\BibitemShut {Stop}%
\bibitem [{\citenamefont {Dodin}\ \emph {et~al.}(2004)\citenamefont {Dodin},
  \citenamefont {Fisch},\ and\ \citenamefont
  {Rax}}]{dodinPonderomotiveBarrierMaxwell2004}%
  \BibitemOpen
  \bibfield  {author} {\bibinfo {author} {\bibfnamefont {I.~Y.}\ \bibnamefont
  {Dodin}}, \bibinfo {author} {\bibfnamefont {N.~J.}\ \bibnamefont {Fisch}},\
  and\ \bibinfo {author} {\bibfnamefont {J.~M.}\ \bibnamefont {Rax}},\ }\href
  {https://doi.org/10.1063/1.1787771} {\bibfield  {journal} {\bibinfo
  {journal} {Physics of Plasmas}\ }\textbf {\bibinfo {volume} {11}},\ \bibinfo
  {pages} {5046} (\bibinfo {year} {2004})}\BibitemShut {NoStop}%
\bibitem [{\citenamefont {Brillouin}(1945)}]{brillouinTheoremLarmorIts1945}%
  \BibitemOpen
  \bibfield  {author} {\bibinfo {author} {\bibfnamefont {L.}~\bibnamefont
  {Brillouin}},\ }\href {https://doi.org/10.1103/PhysRev.67.260} {\bibfield
  {journal} {\bibinfo  {journal} {Physical Review}\ }\textbf {\bibinfo {volume}
  {67}},\ \bibinfo {pages} {260} (\bibinfo {year} {1945})}\BibitemShut
  {NoStop}%
\bibitem [{\citenamefont
  {Davidson}(1990)}]{davidsonPhysicsNonneutralPlasmas1990}%
  \BibitemOpen
  \bibfield  {author} {\bibinfo {author} {\bibfnamefont {R.~C.}\ \bibnamefont
  {Davidson}},\ }\href@noop {} {\emph {\bibinfo {title} {Physics of
  {{Nonneutral Plasmas Addison-Wesley}}, {{Red-wood City}}}}}\ (\bibinfo
  {publisher} {{California}},\ \bibinfo {year} {1990})\BibitemShut {NoStop}%
\bibitem [{\citenamefont {Rax}\ and\ \citenamefont
  {Gueroult}(2016)}]{raxRotationInstabilitiesIsotope2016}%
  \BibitemOpen
  \bibfield  {author} {\bibinfo {author} {\bibfnamefont {J.-M.}\ \bibnamefont
  {Rax}}\ and\ \bibinfo {author} {\bibfnamefont {R.}~\bibnamefont {Gueroult}},\
  }\href {https://doi.org/10.1017/S0022377816000878} {\bibfield  {journal}
  {\bibinfo  {journal} {J. Plasma Phys.}\ }\textbf {\bibinfo {volume} {82}},\
  \bibinfo {pages} {595820504} (\bibinfo {year} {2016})}\BibitemShut {NoStop}%
\bibitem [{\citenamefont
  {Halbach}(1980)}]{halbachDesignPermanentMultipole1980}%
  \BibitemOpen
  \bibfield  {author} {\bibinfo {author} {\bibfnamefont {K.}~\bibnamefont
  {Halbach}},\ }\href {https://doi.org/10.1016/0029-554X(80)90094-4} {\bibfield
   {journal} {\bibinfo  {journal} {Nuclear Instruments and Methods}\ }\textbf
  {\bibinfo {volume} {169}},\ \bibinfo {pages} {1} (\bibinfo {year}
  {1980})}\BibitemShut {NoStop}%
\bibitem [{\citenamefont {Rax}\ \emph {et~al.}(2018)\citenamefont {Rax},
  \citenamefont {Robiche}, \citenamefont {Gueroult},\ and\ \citenamefont
  {Ehrlacher}}]{raxKineticTheoryTransport2018}%
  \BibitemOpen
  \bibfield  {author} {\bibinfo {author} {\bibfnamefont {J.~M.}\ \bibnamefont
  {Rax}}, \bibinfo {author} {\bibfnamefont {J.}~\bibnamefont {Robiche}},
  \bibinfo {author} {\bibfnamefont {R.}~\bibnamefont {Gueroult}},\ and\
  \bibinfo {author} {\bibfnamefont {C.}~\bibnamefont {Ehrlacher}},\ }\href
  {https://doi.org/10.1063/1.5030536} {\bibfield  {journal} {\bibinfo
  {journal} {Physics of Plasmas}\ }\textbf {\bibinfo {volume} {25}},\ \bibinfo
  {pages} {072503} (\bibinfo {year} {2018})}\BibitemShut {NoStop}%
\bibitem [{\citenamefont {Lichtenberg}\ and\ \citenamefont
  {Lieberman}(1983)}]{lichtenbergRegularStochasticMotion1983a}%
  \BibitemOpen
  \bibfield  {author} {\bibinfo {author} {\bibfnamefont {A.~J.}\ \bibnamefont
  {Lichtenberg}}\ and\ \bibinfo {author} {\bibfnamefont {M.~A.}\ \bibnamefont
  {Lieberman}},\ }\href {https://doi.org/10.1007/978-1-4757-4257-2} {\emph
  {\bibinfo {title} {Regular and {{Stochastic Motion}}}}},\ edited by\ \bibinfo
  {editor} {\bibfnamefont {F.}~\bibnamefont {John}}, \bibinfo {editor}
  {\bibfnamefont {J.~E.}\ \bibnamefont {Marsden}},\ and\ \bibinfo {editor}
  {\bibfnamefont {L.}~\bibnamefont {Sirovich}},\ \bibinfo {series} {Applied
  {{Mathematical Sciences}}}, Vol.~\bibinfo {volume} {38}\ (\bibinfo
  {publisher} {{Springer New York}},\ \bibinfo {address} {{New York, NY}},\
  \bibinfo {year} {1983})\BibitemShut {NoStop}%
\bibitem [{\citenamefont {Boris}(1970)}]{boris1970relativistic}%
  \BibitemOpen
  \bibfield  {author} {\bibinfo {author} {\bibfnamefont {J.~P.}\ \bibnamefont
  {Boris}},\ }in\ \href@noop {} {\emph {\bibinfo {booktitle} {Proc. {{Fourth}}
  Conf. {{Num}}. {{Sim}}. {{Plasmas}}}}}\ (\bibinfo {year} {1970})\ pp.\
  \bibinfo {pages} {3--67}\BibitemShut {NoStop}%
\bibitem [{\citenamefont {Stoltz}\ \emph {et~al.}(2002)\citenamefont {Stoltz},
  \citenamefont {Cary}, \citenamefont {Penn},\ and\ \citenamefont
  {Wurtele}}]{stoltzEfficiencyBorislikeIntegration2002}%
  \BibitemOpen
  \bibfield  {author} {\bibinfo {author} {\bibfnamefont {P.~H.}\ \bibnamefont
  {Stoltz}}, \bibinfo {author} {\bibfnamefont {J.~R.}\ \bibnamefont {Cary}},
  \bibinfo {author} {\bibfnamefont {G.}~\bibnamefont {Penn}},\ and\ \bibinfo
  {author} {\bibfnamefont {J.}~\bibnamefont {Wurtele}},\ }\href
  {https://doi.org/10.1103/PhysRevSTAB.5.094001} {\bibfield  {journal}
  {\bibinfo  {journal} {Physical Review Special Topics - Accelerators and
  Beams}\ }\textbf {\bibinfo {volume} {5}},\ \bibinfo {pages} {094001}
  (\bibinfo {year} {2002})}\BibitemShut {NoStop}%
\bibitem [{\citenamefont {Qin}\ \emph {et~al.}(2013)\citenamefont {Qin},
  \citenamefont {Zhang}, \citenamefont {Xiao}, \citenamefont {Liu},
  \citenamefont {Sun},\ and\ \citenamefont
  {Tang}}]{qinWhyRelaxBorisAlgorithm2013}%
  \BibitemOpen
  \bibfield  {author} {\bibinfo {author} {\bibfnamefont {H.}~\bibnamefont
  {Qin}}, \bibinfo {author} {\bibfnamefont {S.}~\bibnamefont {Zhang}}, \bibinfo
  {author} {\bibfnamefont {J.}~\bibnamefont {Xiao}}, \bibinfo {author}
  {\bibfnamefont {J.}~\bibnamefont {Liu}}, \bibinfo {author} {\bibfnamefont
  {Y.}~\bibnamefont {Sun}},\ and\ \bibinfo {author} {\bibfnamefont {W.~M.}\
  \bibnamefont {Tang}},\ }\href {https://doi.org/10.1063/1.4818428} {\bibfield
  {journal} {\bibinfo  {journal} {Phys. Plasmas}\ }\textbf {\bibinfo {volume}
  {20}},\ \bibinfo {pages} {084503} (\bibinfo {year} {2013})}\BibitemShut
  {NoStop}%
\bibitem [{\citenamefont {Ochs}\ and\ \citenamefont
  {Fisch}(2021)}]{ochsNonresonantDiffusionAlpha2021a}%
  \BibitemOpen
  \bibfield  {author} {\bibinfo {author} {\bibfnamefont {I.~E.}\ \bibnamefont
  {Ochs}}\ and\ \bibinfo {author} {\bibfnamefont {N.~J.}\ \bibnamefont
  {Fisch}},\ }\href {https://doi.org/10.1103/PhysRevLett.127.025003} {\bibfield
   {journal} {\bibinfo  {journal} {Physical Review Letters}\ }\textbf {\bibinfo
  {volume} {127}},\ \bibinfo {pages} {025003} (\bibinfo {year}
  {2021})}\BibitemShut {NoStop}%
\bibitem [{\citenamefont {Ochs}\ and\ \citenamefont
  {Fisch}(2023)}]{ochsPonderomotiveRecoilElectromagnetic2023}%
  \BibitemOpen
  \bibfield  {author} {\bibinfo {author} {\bibfnamefont {I.~E.}\ \bibnamefont
  {Ochs}}\ and\ \bibinfo {author} {\bibfnamefont {N.~J.}\ \bibnamefont
  {Fisch}},\ }\href {https://doi.org/10.1063/5.0138384} {\bibfield  {journal}
  {\bibinfo  {journal} {Physics of Plasmas}\ }\textbf {\bibinfo {volume}
  {30}},\ \bibinfo {pages} {022102} (\bibinfo {year} {2023})}\BibitemShut
  {NoStop}%
\end{thebibliography}%
\end{document}